\documentclass[a4paper,fleqn]{cas-sc}
\usepackage[square,numbers,sort&compress]{natbib}
\usepackage[capitalise]{cleveref}
\usepackage{mathrsfs} 
\usepackage[figurename=Fig.]{caption}
\usepackage{subfig}
\usepackage{todonotes}

\ExplSyntaxOn              
\keys_set:nn { stm / mktitle } { nologo }   
\ExplSyntaxOff    

\def\tsc#1{\csdef{#1}{\textsc{\lowercase{#1}}\xspace}}
\tsc{WGM}
\tsc{QE}
\tsc{EP}
\tsc{PMS}
\tsc{BEC}
\tsc{DE}

\begin{document}


\let\WriteBookmarks\relax
\def\floatpagepagefraction{1}
\def\textpagefraction{.001}

\shorttitle{XOSBPD}

\shortauthors{Q. Liu et~al.}

\title [mode = title]{An extended ordinary state-based peridynamics for non-spherical horizons}                      



\author[1]{Qibang Liu}[type=editor,
                        auid=000,bioid=1,
                        orcid=0000-0001-7935-7907]

\cormark[1]
\ead{qibangliu@ksu.edu}




\credit{Conceptualization, Methodology, Software, Original draft}

\author[2]{Muhao Chen}[type=editor,
auid=000,bioid=1,
orcid=0000-0003-1812-6835]


\author[2]{Robert E. Skelton}[type=editor,
auid=000,bioid=1,
orcid=0000-0001-6503-9115]


\address[1]{Department of Mechanical and Nuclear Engineering, Kansas State University, Manhattan, KS, 66506, USA}
\address[2]{Department of Petroleum Engineering, Texas A\&M University, College Station, TX,77840, USA}

\cortext[cor1]{Corresponding author}

\begin{abstract}
This work presents an extended ordinary state-based peridynamics (XOSBPD) model for the non-spherical horizons. Based on the OSBPD, we derive the XOSBPD by introducing the Lagrange multipliers to guarantee the non-local dilatation and non-local strain energy density (SED) are equal to local dilatation and local SED, respectively. In this formulation, the XOSBPD removes the limitation of spherical horizons and is suitable for arbitrary horizon shapes. In addition, the presented XOSBPD does not need volume and surface correction and allows non-uniform discretization implementation with various horizon sizes. Three classic examples demonstrate the accuracy and capability for complex dynamical fracture analysis. The proposed method provides an efficient tool and in-depth insight into the failure mechanism of structure components and solid materials. 
\end{abstract}



\begin{graphicalabstract}
\includegraphics[width=135mm]{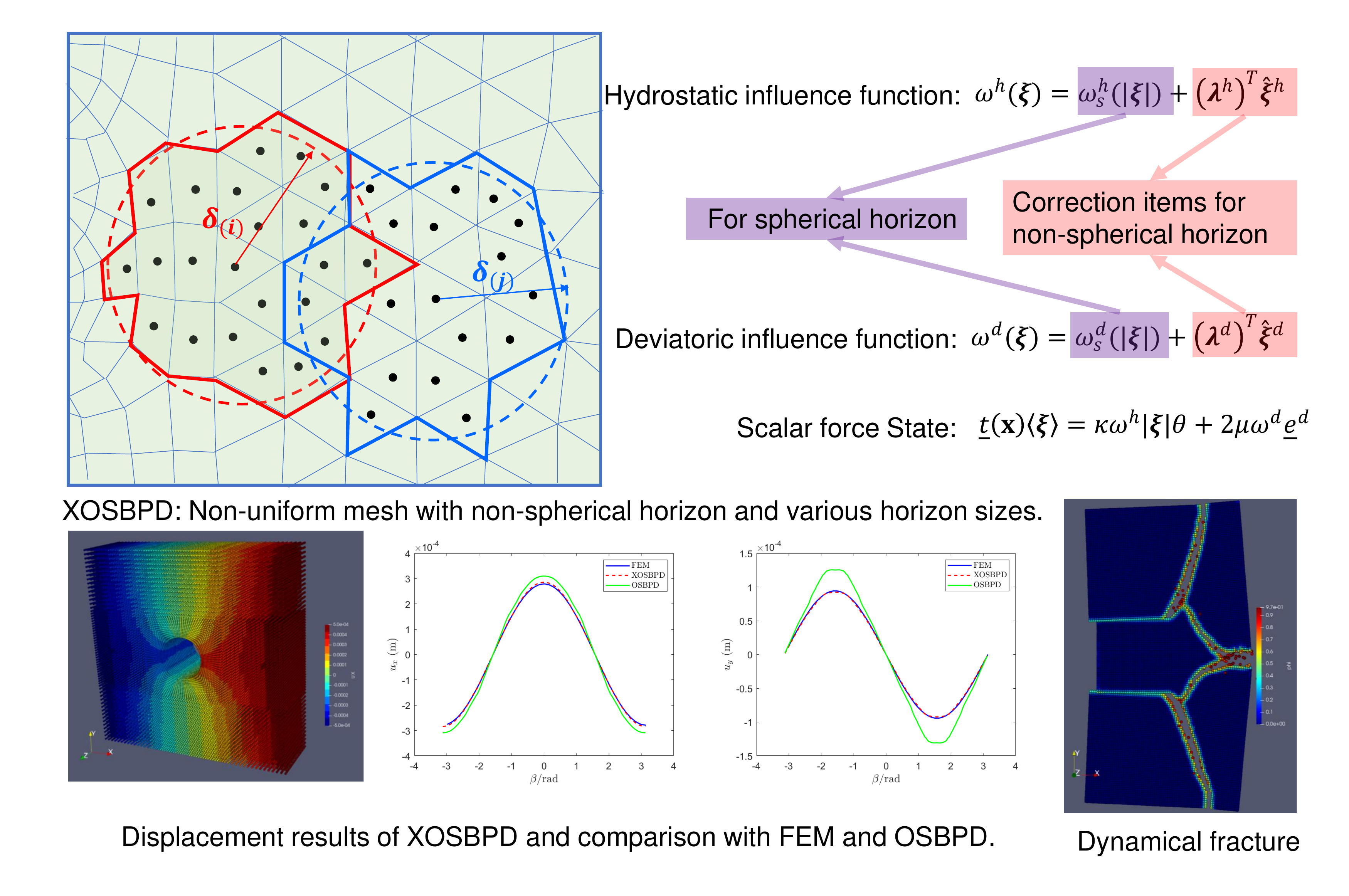}
\end{graphicalabstract}

\begin{highlights}
\item An extended ordinary state-based peridynamics (XOSBPD) is proposed.
\item XOSBPD extends the horizon from spherical to non-spherical ones.
\item XOSBPD removes the requirement of volume and surface corrections.
\item XOSBPD enables the implementation of non-uniform mesh with various horizon sizes.
\end{highlights}


\maketitle

\section{Introduction}

The failure mechanism of structure components and solid materials is fundamental for the research of structural integrity. Classical continuum mechanics (CCM) employs spatial derivatives to describe the material behaviors with a requirement of continuum material properties during the deformation. However, derivatives are undefined at discontinuities of materials. Thus the CCM is inherently difficult to material failure. In 2000, the peridynamics (PD) theory \citep{silling2000reformulation} was introduced to remove this drawback. The critical idea of the PD theory is that the PD employs spatial integrals rather than derivatives. Since the integrals are still defined at discontinuities, the PD is suitable for describing non-continuum materials.


The first proposed PD theory is the bond-based formulation \citep{silling2000reformulation} which treats the bond between two material points as a spring, resulting in a restriction of a fixed Poisson's ratio of 1/4 \citep{gerstle2005peridynamic,madenci_peridynamic_2014}. This substantial limitation blocks the PD from a broader application with various materials. Thus, the fixed Poisson's ratio requirement must be removed. Indeed, \citet{silling_peridynamic_2007} extended the PD and developed the state-based PD (SBPD) to eliminate this constraint. In the SBPD, the bond force is redefined not only by the bond between the two material points but also by other neighboring bonds. \citet{silling_peridynamic_2007} provided two versions of the SBPD: the ordinary state-based PD (OSBPD) and the non-ordinary state-based PD (NOSBPD). The OSBPD divided the stretch of a bond into a hydrostatic and a deviatoric part, similar to dividing the strain tensor into hydrostatic and deviatoric strain tensors. The NOSBPD introduced a non-local displacement gradient and employed classical constitutive law to describe the bond forces. Both OSBPD and NOSBPD can remove the restriction of Poisson's ratio. Note that NOSBPD is inherent unstable, thus requires zero-energy mode control \cite{breitenfeld_non-ordinary_2014}. Thus, in this paper, an extended OSBPD is developed based on the OSBPD model.


Although the SBPD provides a theoretical framework for the discontinuity analysis with various materials, there are still three major issues that attract researchers to improve the generosity and decrease the computational cost: (a) the surface correction, (b) the volume correction, and (c) the meshless method with uniform discretization. All three problems are related to the spherical horizons in the PD. The PD horizon is defined as an interaction domain where the classical strain energy equals that of a PD. The requirement of a spherical horizon might be reasonable in the inner areas away from the structure boundaries. Nevertheless, as we get close to the edges of the structure, the spherical horizons are usually truncated, called the surface effect in PD. Several methods have been proposed to reduce or eliminate the surface effect. For example, \citet{gerstle2005peridynamic} and \citet{oterkus_peridynamic_2014} introduced fictitious nodes outsides of the boundary which may only be applied for simple geometries. \citet{scabbia2021novel} proposed a revised fictitious node approach with an extrapolation by Taylor series expansion to reduce the surface effect. \citet{madenci_peridynamic_2014} computed approximate corrections of the material constants for nodes near the surface by equating local SED to the non-local SED. This approach is exact in only homogeneous deformations because it assumes six simple loading conditions applied on the whole domain for surface correction:
\begin{subequations}\label{eq:loadcondi}
\begin{align}
    \varepsilon_{11}\neq 0,\ \varepsilon_{22}=\varepsilon_{33}=\varepsilon_{12}=\varepsilon_{23}=\varepsilon_{31}=0,\\
     \varepsilon_{22}\neq 0,\ \varepsilon_{11}=\varepsilon_{33}=\varepsilon_{12}=\varepsilon_{23}=\varepsilon_{31}=0,\\
      \varepsilon_{33}\neq 0,\ \varepsilon_{11}=\varepsilon_{22}=\varepsilon_{12}=\varepsilon_{23}=\varepsilon_{31}=0,\\
       \varepsilon_{12}\neq 0,\ \varepsilon_{11}=\varepsilon_{22}=\varepsilon_{33}=\varepsilon_{23}=\varepsilon_{31}=0,\\
        \varepsilon_{23}\neq 0,\ \varepsilon_{11}=\varepsilon_{22}=\varepsilon_{33}=\varepsilon_{12}=\varepsilon_{31}=0,\\
         \varepsilon_{31}\neq 0,\ \varepsilon_{11}=\varepsilon_{22}=\varepsilon_{33}=\varepsilon_{12}=\varepsilon_{23}=0.
\end{align}
\end{subequations}
\citet{shen_peridynamic_2021} proposed a similar surface correction method yet requires the FEM simulation performed forward in detail.


Moreover, the numerical implementation of the PD is generally by a simple meshfree method with a one-point integration method \citep{silling_meshfree_2005}. Such quadrature requires a volume correction to improve the integration accuracy. The simplest volume correction method is introducing a linear correction factor \citep{silling_meshfree_2005,parks_implementing_2008,le2014two}. Some other more accurate correction methods can be found in \citep{yu_new_2011,seleson_improved_2014,ren_dual-horizon_2017,seleson_numerical_2018}, but usually with a higher computational cost. 

Furthermore, because of the spherical horizon restriction, one has to use regular uniform numerical discretization with constant horizon size to solve the PD equations instead of irregular non-uniform meshes with various horizon sizes. Otherwise, one may run into a "ghost force" problem which breaks the conservation of linear and angular momentum. However, it is also clear that uniform discretization has two drawbacks: (a). A high computational cost because the refinement level is applied anywhere in the domain, (b). The uniform mesh is difficult to conformal to complex geometries. To solve the "ghost force" issue and to allow numerical implementation of non-uniform discretization with various horizon sizes, \citet{ren_dual-horizon_2016,ren_dual-horizon_2017} proposed a dual-horizon PD model.


To solve the three critical problems of volume and surface corrections and uniform discretization implementation induced by the spherical horizon as a whole, many researchers have proposed a few PD models with non-spherical horizons. For example, \citet{madenci_peridynamic_2019} developed a PD-least square minimization (PDLSM), and \citet{liu_revised_2021} formulated a revised NOSBPD based on Taylor series expansion and least square minimization, which works for arbitrary horizon shapes. Still, there are instability problems in the dynamical analysis of the PDLSM and NOSBPD. \citet{madenci_bond-based_2021} derived bond-based peridynamics (BBPD) with stretch and rotation Kinematics based on the PDLSM. \citet{hu_thermomechanical_2018} presented a generalized BBPD and OSBPD models with non-spherical horizons, where they also assumed a simple loading condition applied to the whole domain to approximately correct material constants, i.e. $\varepsilon_{11}=\varepsilon_{22}=\varepsilon_{33}\neq 0$ and $\varepsilon_{12}=\varepsilon_{23}=\varepsilon_{31}=0$. \citet{mitchell_position-aware_2015} demonstrated a position-aware OSBPD model for non-spherical horizons in which the six simple loading conditions of \cref{eq:loadcondi} were assumed for correcting the scalar force states. Here, we must point out that the simple homogeneous deformation assumptions may bring errors because the real domain solutions are generally complex. Thus, an SBPD model is still needed without the assumptions given by \cref{eq:loadcondi}.


Based on the OSBPD, this paper presents an extended OSBPD (XOSBPD) model for non-spherical horizons by introducing the Lagrange multipliers to guarantee the non-local dilatation and non-local strain energy density (SED) are equal to local dilatation and local SED, respectively. The significant advantages of the XOSBPD are: (a). the XOSBPD does not assume the simple loading condition given in \cref{eq:loadcondi} and works for arbitrary deformation conditions, (b). no surface or volume corrections are needed, and (c). the XOSBPD allows non-uniform discretization implementation with various horizon sizes.

This paper is organized as follows. The OSBPD theory is briefly reviewed in \cref{sec:OSBPD}, and the XOSBPD is proposed and presented in \cref{sec:XOSBPD}. After that, the numerical implementation of XOSBPD is described in \cref{sec:NumImp}. Then, three classic examples are performed to demonstrate the proposed XOSBPD in \cref{sec:NumR}. Finally, the conclusions are drawn in \cref{sec:conc}.


\section{Ordinary state based PD}\label{sec:OSBPD}



In this section, the OSBPD for the 3-D domain first developed by \citet{silling_peridynamic_2007} is briefly reviewed for completeness. The equation of motion in the PD is given by:
\begin{equation}\label{eq:eqm}
    \rho \Ddot{\textbf{u}}=\textbf{L}(\textbf{x},t)+\textbf{b}(\textbf{x},t),
\end{equation}
where $\rho$ is the material mass density, $\textbf{u}$ is the displacement vector field, $\textbf{L}$ is non-local internal force density, and $\textbf{b}$ is externally applied body force density. The $\textbf{L}$ can be evaluated as:
\begin{equation}\label{eq:sillL}
   \textbf{L}(\textbf{x},t)=\int_{H_x} \left\{ \underline{\textbf{T}}(\textbf{x}) \langle \boldsymbol{\xi} \rangle - \underline{\textbf{T}}(\textbf{x}') \langle -\boldsymbol{\xi} \rangle \right\} \mathrm{d} V_{x'},
\end{equation}
where $H_x$ is the horizon of point $\textbf{x}$ and is required to be a sphere for both BBPD and SBPD \cite{silling2000reformulation,silling_peridynamic_2007}. $\boldsymbol{\xi}$ is a bond vector defined as:
\begin{equation}
    \boldsymbol{\xi}=\textbf{x}'-\textbf{x},
\end{equation}
and $\underline{\textbf{T}}(\textbf{x}) \langle \boldsymbol{\xi} \rangle$ is the force state. The \emph{state} is a mathematical object represented with a underline which maps the bond vector $\boldsymbol{\xi}$ to a tensor. For example, the un-deformed bond length state $ \underline{x}$ is defined as:
\begin{equation}
    \underline{x}\langle \boldsymbol{\xi} \rangle=|\boldsymbol{\xi}|.
\end{equation}
And the deformation vector state $\underline{\boldsymbol{Y}}$ is defined as:
\begin{equation}
    \underline{\boldsymbol{Y}}\langle \boldsymbol{\xi} \rangle=\boldsymbol{\xi}+\boldsymbol{\eta},
\end{equation}
in which $\boldsymbol{\eta}=\textbf{u}\left(\textbf{x}'\right)-\textbf{u}\left(\textbf{x}\right)$ is the relative displacement of the bond. One important scalar state in OSBPD is the extension state $\underline{e}$ defined as:
\begin{equation}
    \underline{e}\langle \boldsymbol{\xi} \rangle=|\underline{\boldsymbol{Y}}|-\underline{x}.
\end{equation}
Similar to dividing the strain into the hydrostatic and deviatoric parts in the CCM, the extension state can also divided into two parts:
\begin{equation}
    \underline{e}=\underline{e}^h+\underline{e}^d,
\end{equation}
in which the superscripts $h$ and $d$ represent the hydrostatic and deviatoric, respectively. The hydrostatic extension state is defined as:
\begin{equation}
    \underline{e}^h= \frac{\theta \underline{x}}{ 3},
\end{equation}
where $\theta$ is the non-local dilatation and for a spherical horizon:
\begin{equation}\label{eq:theta_sph}
    \theta =\theta \left(\underline{e} \right) =\frac{3}{m}\left(\underline{\omega} \underline{x} \right)\bullet \underline{e},
\end{equation}
where $\underline{\omega}\langle \boldsymbol{\xi} \rangle=\omega\left(|\boldsymbol{\xi}| \right)$ is an influence function and $m$ is the weighted volume, defined as:
\begin{equation}
  m=  \left(\underline{\omega} \underline{x} \right)\bullet \underline{x}.
\end{equation}
Here, the dot product of two scalar states of $\underline{a}$ and $\underline{b}$ is defined as: 
\begin{equation}
    \underline{a}  \bullet \underline{b}=\int_{H_x} \underline{a}\langle \boldsymbol{\xi} \rangle \underline{b}\langle \boldsymbol{\xi} \rangle \mathrm{d} V_{x'}.
\end{equation}
For example, the weighted volume is evaluated as:
\begin{equation}
    m=\int_{H_x} \omega\left(|\boldsymbol{\xi}| \right) |\boldsymbol{\xi}|^2 \mathrm{d} V_{x'}=\int_{0}^{\delta} \int^{2\pi}_0 \int^{\pi}_0  \omega(r) r^4 \sin \phi \mathrm{d} \phi \mathrm{d} \theta \mathrm{d}r=4\pi \int_{0}^{\delta} \omega(r)r^4 \mathrm{d}r,
\end{equation}
where $\delta$ is the horizon size.

For the OSBPD, the force vector states $\underline{\textbf{T}}(\textbf{x}) \langle \boldsymbol{\xi} \rangle$ is defined as:
\begin{equation}\label{eq:T}
    \underline{\textbf{T}}(\textbf{x}) \langle \boldsymbol{\xi} \rangle=\underline{t}(\textbf{x}) \langle \boldsymbol{\xi} \rangle \frac{\boldsymbol{\eta}+\boldsymbol{\xi}}{|\boldsymbol{\eta}+\boldsymbol{\xi}|},
\end{equation}
where $\underline{t}(\textbf{x}) \langle \boldsymbol{\xi} \rangle$ is the scalar force state and is derived from the Fréchet derivative of the strain energy density with respect to the extension state as: 
\begin{equation}\label{eq:frederi}
    \underline{t}(\textbf{x}) \langle \boldsymbol{\xi} \rangle =\left(\frac{\partial W_{pd}}{\partial \underline{e}} \right)^f =\left(\frac{\partial W_{pd}}{\partial \underline{e}^h} \right)^f + \left(\frac{\partial W_{pd}}{\partial \underline{e}^d} \right)^f,
\end{equation}
in which the superscript $f$ represents the Fréchet derivative is defined in \citep{silling_peridynamic_2007}. 
For an isotropic material and a spherical horizon, the non-local strain energy density can be expressed as:
\begin{equation}\label{eq:sed_sph}
    W_{pd}=\frac{\kappa \theta^2}{2}+\frac{15}{2m}\mu(\underline{\omega} \underline{e}^d )\bullet \underline{e}^d,
\end{equation}
in which $\kappa$ is the bulk modulus and $\mu$ is the shear modulus. Substituting \cref{eq:sed_sph} into \cref{eq:frederi}, we have the scalar force state:
\begin{equation}\label{eq:finscalfor_sph}
     \underline{t}(\textbf{x}) \langle \boldsymbol{\xi} \rangle=\frac{3 \kappa \theta}{m} \underline{\omega} \underline{x} + \frac{15}{m} \mu \underline{\omega} \underline{e}^d.
\end{equation}
It is worth noting that \cref{eq:theta_sph,eq:sed_sph} are only valid for the spherical horizon. Thus, \cref{eq:finscalfor_sph} only works for the spherical horizon.  

\section{Extended ordinary state based PD}\label{sec:XOSBPD}

In this section, we extend the horizons of the OSBPD from sphere to non-sphere ones. Thus removing the requirements of volume and surface corrections and enabling non-uniform discretization implementation with various horizon sizes. The presented XOSBPD introduces two correction items for the non-spherical horizon and arbitrary deformation conditions.

As described in \cref{sec:OSBPD}, the essential idea of the OSBPD are: (a) dividing the extension state $\underline{e}$ into hydrostatic extension state $\underline{e}^h$ and deviatoric extension state $\underline{e}^d$, (b) expressing the non-local strain energy density in terms of the $\underline{e}^h$ and $\underline{e}^d$, and then (c) using \cref{eq:frederi} to derive the scalar force state $\underline{t}$. The splitting of the extension state $\underline{e}$ is based on the assumption of small deformation and the classical kinematics as:
\begin{equation}
\begin{split}
     \underline{e}&=|\underline{\boldsymbol{Y}}|-\underline{x}=\frac{\boldsymbol{\xi}\cdot \boldsymbol{\varepsilon}\boldsymbol{\xi}}{|\boldsymbol{\xi}|}\\
     &=\frac{\boldsymbol{\xi}\cdot \boldsymbol{\varepsilon}^h\boldsymbol{\xi}}{|\boldsymbol{\xi}|} + \frac{\boldsymbol{\xi}\cdot \boldsymbol{\varepsilon}^d\boldsymbol{\xi}}{|\boldsymbol{\xi}|}\\
     &=\underline{e}^h+\underline{e}^d,
\end{split}
\end{equation}
in which $\boldsymbol{\varepsilon}$, $\boldsymbol{\varepsilon}^h$, and $\boldsymbol{\varepsilon}^d$ are the strain tensor, hydrostatic strain tensor, and deviatoric strain tensor, respectively. Note that the non-local dilatation must equal to the local dilatation, that is $\theta=\varepsilon_{ii}$. Thus, we have:
\begin{align}\label{eq:eh_xpd}
    \underline{e}^h & =\frac{\boldsymbol{\xi}\cdot \boldsymbol{\varepsilon}^h\boldsymbol{\xi}}{|\boldsymbol{\xi}|}=\frac{\boldsymbol{\xi}\cdot \frac{\theta \textbf{I}}{n_d}\boldsymbol{\xi}}{|\boldsymbol{\xi}|}=\frac{\theta \underline{x}}{n_d},\\
    \label{eq:ed_xpd}
    \underline{e}^d & =\frac{\boldsymbol{\xi}\cdot \boldsymbol{\varepsilon}^d\boldsymbol{\xi}}{|\boldsymbol{\xi}|} =\underline{e}- \frac{\theta \underline{x}}{n_d},
\end{align}
in which $n_d$ is the number of dimensions.
From the CCM, the strain energy density of an isotropic material is:
\begin{equation}\label{eq:sde_ccm}
    W_{CCM}= \frac{1}{2} \kappa \varepsilon_{ii}\varepsilon_{jj} + \mu \boldsymbol{\varepsilon}^d_{ij}\boldsymbol{\varepsilon}^d_{ij}.
\end{equation}
As stated in \cite{silling_peridynamic_2007}, the non-local SED defined in \cref{eq:sed_sph} equals to that of the local SED only for the spherical horizon. To extend to the OSBPD for non-spherical horizons, we redefined the non-local strain energy density as:
\begin{equation}\label{eq:sde_xpd}
     W_{pd}=\frac{\kappa \theta^2}{2}+\mu(\underline{\omega}^d \underline{e}^d )\bullet \underline{e}^d.
\end{equation}
In this study, we refer the $\underline{\omega}^d\langle \boldsymbol{\xi} \rangle=\omega^d( \boldsymbol{\xi})$ as deviatoric influence function. For spherical horizon, ${\omega}^d$ is defined as:
\begin{equation}
   {\omega}^d={\omega}^d_s,
\end{equation}
in which ${\omega}^d_s$ is the spherical deviatoric influence function, and comparing \cref{eq:sed_sph} with \cref{eq:sde_xpd} reveals that the ${\omega}^d_s$ is:
\begin{equation}
    {\omega}^d_s=\frac{n_d(n_d+2)}{2m}\omega\left(|\boldsymbol{\xi}| \right),
\end{equation}
where the weighted volume $m$ is defined as:
\begin{equation}
    m=
    \begin{cases}
     &4\pi \int_{0}^{\delta} \omega(r)r^4 \mathrm{d}r, \quad \text{3-D},\\
     &2\pi \int_{0}^{\delta} \omega(r)r^3 \mathrm{d}r, \quad \text{2-D}.
    \end{cases}
\end{equation}
Besides, the definition of the non-local dilatation in \cref{eq:theta_sph} can ensure $\theta=\varepsilon_{ii}$ only for spherical horizon. For non-spherical horizon, we redefine the non-local dilatation as below:
\begin{equation}\label{eq:theta_xpd}
    \theta=(\underline{\omega}^h \underline{x}) \bullet \underline{e}.
\end{equation}
In this work, we refer the $\underline{\omega}^h\langle \boldsymbol{\xi} \rangle=\omega^h( \boldsymbol{\xi})$ as hydrostatic influence function. For spherical horizon, ${\omega}^h$ is defined as:
\begin{equation}
     {\omega}^h={\omega}^h_s,
\end{equation}
in which ${\omega}^h_s$ is the spherical hydrostatic influence function. Comparing \cref{eq:theta_sph} and \cref{eq:theta_xpd} reveals that the ${\omega}^h_s$ is expressed as:
\begin{equation}
    {\omega}^h_s=\frac{n_d}{m}\omega\left(|\boldsymbol{\xi}| \right).
\end{equation}
Equating \cref{eq:sde_xpd} to \cref{eq:sde_ccm} leads to:
\begin{equation}\label{eq:condi}
    \theta=\varepsilon_{ii}\quad  \mathrm{and}\quad (\underline{\omega}^d \underline{e}^d )\bullet \underline{e}^d=\boldsymbol{\varepsilon}^d:\boldsymbol{\varepsilon}^d.
\end{equation}
In the following section, we derive the hydrostatic and the deviatoric influence functions, satisfying the \cref{eq:condi} for arbitrary horizon shapes.

\subsection{Hydrostatic influence function}

Consider a 3-D body ($n_d=3$), we rearrange $\theta=\varepsilon_{ii}$ as:
\begin{equation}\label{eq:derivOmegaH}
\begin{split}
   0&=(\underline{\omega}^h \underline{x}) \bullet \underline{e}- \left(\varepsilon_{11}+\varepsilon_{22}+\varepsilon_{33}\right)\\
   &=\int_{H_x} {\omega}^h |\boldsymbol{\xi}| e \mathrm{d}V_{x'} - \left(\varepsilon_{11}+\varepsilon_{22}+\varepsilon_{33}\right)\\
    &=\int_{H_x} {\omega}^h |\boldsymbol{\xi}| \frac{\boldsymbol{\xi}\cdot \boldsymbol{\varepsilon}\boldsymbol{\xi}}{|\boldsymbol{\xi}|} \mathrm{d}V_{x'}- \left(\varepsilon_{11}+\varepsilon_{22}+\varepsilon_{33}\right)\\
    &=\varepsilon_{11}\left( \int_{H_x} {\omega}^h \xi_1^2 \mathrm{d}V_{x'} -1 \right) + \varepsilon_{22}\left(\int_{H_x} {\omega}^h \xi_2^2\mathrm{d}V_{x'} -1 \right) + \varepsilon_{33}\left(\int_{H_x} {\omega}^h \xi_3^2\mathrm{d}V_{x'} -1 \right)\\
    &+\varepsilon_{12}\int_{H_x} {\omega}^h \xi_1\xi_2\mathrm{d}V_{x'} + \varepsilon_{23}\int_{H_x} {\omega}^h \xi_2\xi_3\mathrm{d}V_{x'} + \varepsilon_{31}\int_{H_x} {\omega}^h \xi_3\xi_1\mathrm{d}V_{x'}.
\end{split}
\end{equation}
For an arbitrary strain tensor $\boldsymbol{\varepsilon}$, \cref{eq:derivOmegaH} requires:
\begin{equation}\label{eq:hrocondi}
h_k =0,~ (k=1,2, \cdots, 6),
\end{equation}
in which $h_k$ are defined as:
\begin{subequations}
\label{eq:hyroCond}
\begin{align}
  \label{eq:hyroCond_a}
  &h_1=\int_{H_x} {\omega}^h \xi_1^2 \mathrm{d}V_{x'}-1, \\
  \label{eq:hyroCond_b}
  &h_2=\int_{H_x} {\omega}^h \xi_2^2 \mathrm{d}V_{x'}-1, \\
    \label{eq:hyroCond_c}
 &h_3=\int_{H_x} {\omega}^h \xi_3^2 \mathrm{d}V_{x'}-1, \\
    \label{eq:hyroCond_d}
  &h_4=\int_{H_x} {\omega}^h \xi_1\xi_2 \mathrm{d}V_{x'}, \\
    \label{eq:hyroCond_e}
  &h_5=\int_{H_x} {\omega}^h \xi_2\xi_3 \mathrm{d}V_{x'}, \\
    \label{eq:hyroCond_f}
  &h_6=\int_{H_x} {\omega}^h \xi_3\xi_1 \mathrm{d}V_{x'}. 
\end{align}
\end{subequations}
To find the ${\omega}^h$ to satisfy \cref{eq:hrocondi} for arbitrary shapes of horizon $H_x$, we defined a function $I$ as:
\begin{equation}
    I(\omega^h,\lambda_1,\cdots, \lambda_6)=\frac{1}{2}\left({\omega^h-\omega^h_s} \right) \bullet \left({\omega^h-\omega^h_s} \right)- \sum_{k=1}^6 \lambda^h_k h_k,
\end{equation}
in which $\lambda^h_k$ are the Lagrange multipliers. The first variation of $I$ is:
\begin{equation}
\delta I = \left(\frac{\partial I}{\partial \omega^h } \right)^f \bullet \delta \omega^h + \sum_{k=1}^6 \frac{\partial I}{\partial \lambda^h_k} \delta \lambda^h_k,
\end{equation}
in which the superscript $f$ represents the Fréchet derivative and $\delta$ is the variation.
The condition $ \frac{\partial I}{\partial \lambda^h_k}=0$ leads to \cref{eq:hrocondi}.
The condition $\left(\frac{\partial I}{\partial \omega^h } \right)^f=0$ gives:
\begin{equation}\label{eq:omegah}
    \omega^h(\boldsymbol{\xi})=\omega^h_s(|\boldsymbol{\xi}|)+(\boldsymbol{\lambda}^h)^T \hat{\boldsymbol{\xi}}^h,
\end{equation}
in which $(\boldsymbol{\lambda}^h)^T \hat{\boldsymbol{\xi}}^h$ is the correction item for non-spherical horizon. $\boldsymbol{\lambda}^h$ and $\hat{\boldsymbol{\xi}}^h$ are defined as follows:
\begin{align}
    \boldsymbol{\lambda}^h & =\left[
    \begin{matrix}
    \lambda^h_1 & \lambda^h_2 & \lambda^h_3 & \lambda^h_4 & \lambda^h_5 & \lambda^h_6
    \end{matrix}
   \right]^T,\\
     \hat{\boldsymbol{\xi}}^h  & =\left[
    \begin{matrix}
    \xi_1^2 & \xi_2^2 & \xi_3^2 & \xi_1 \xi_2 & \xi_2 \xi_3 & \xi_3 \xi_1
    \end{matrix}
   \right]^T.
\end{align}
Substituting \cref{eq:omegah} into \cref{eq:hrocondi}, we have:
\begin{equation}\label{eq:lambdah}
   \boldsymbol{\lambda}^h=\left( \textbf{A}^h \right)^{-1}\textbf{R}^h,
\end{equation}
in which:
\begin{align}
    \textbf{A}^h & =\int_{H_x} \hat{\boldsymbol{\xi}}^h \otimes \hat{\boldsymbol{\xi}}^h \mathrm{d}V_{x'},\\
    \textbf{R}^h & =\textbf{r}^h- \int_{H_x} \omega_s^h \hat{\boldsymbol{\xi}}^h \mathrm{d}V_{x'},\\
    \textbf{r}^h & =\left[
    \begin{matrix}
    1 & 1 & 1 & 0 & 0 & 0
    \end{matrix}
 \right]^T.
\end{align}
The Lagrange multiplier $ \boldsymbol{\lambda}^h$ can be solved by \cref{eq:lambdah}. Then, the hydrostatic influence function is evaluated by \cref{eq:omegah}, which satisfies $\theta=\varepsilon_{ii}$ for arbitrary horizon shapes. It is worth noting that $\textbf{R}^h=0$, when $H_x$ is a sphere, leads to $\boldsymbol{\lambda}^h=0$ and $\omega^h=\omega^h_s$. Thus, the definition of the non-local dilatation of the XOSBPD in \cref{eq:theta_xpd} recovers that of the OSBPD.

Besides, for 2D problems, the vectors $\boldsymbol{\lambda}^h$, $\hat{\boldsymbol{\xi}}^h$ and $\textbf{r}^h$ are reduced to as:
\begin{align}
    \boldsymbol{\lambda}^h & =\left[
    \begin{matrix}
    \lambda^h_1 & \lambda^h_2 & \lambda^h_3
    \end{matrix}
   \right]^T,\\
    \hat{\boldsymbol{\xi}}^h & =\left[
    \begin{matrix}
    \xi_1^2 & \xi_2^2 & \xi_1 \xi_2
    \end{matrix}
   \right]^T,\\
    \textbf{r}^h & =\left[
    \begin{matrix}
    1 & 1 & 0
    \end{matrix}
 \right]^T.
\end{align} 

\subsection{Deviatoric influence function} 

Consider a 3-D body ($n_d=3$), we rearrange $ (\underline{\omega}^d \underline{e}^d )\bullet \underline{e}^d=\boldsymbol{\varepsilon}^d:\boldsymbol{\varepsilon}^d$ as:
\begin{equation}\label{eq:derivOmegaD1}
\begin{split}
    0&=(\underline{\omega}^d \underline{e}^d )\bullet \underline{e}^d - \boldsymbol{\varepsilon}^d:\boldsymbol{\varepsilon}^d\\
    &=\int_{H_x} {\omega}^d \underline{e}^d \underline{e}^d \mathrm{d}V_{x'} - \boldsymbol{\varepsilon}^d:\boldsymbol{\varepsilon}^d\\
    &=\int_{H_x} {\omega}^d \frac{\boldsymbol{\xi}\cdot \boldsymbol{\varepsilon}^d\boldsymbol{\xi}}{|\boldsymbol{\xi}|} \frac{\boldsymbol{\xi}\cdot \boldsymbol{\varepsilon}^d\boldsymbol{\xi}}{|\boldsymbol{\xi}|} \mathrm{d}V_{x'} - \boldsymbol{\varepsilon}^d:\boldsymbol{\varepsilon}^d\\
    &=\varepsilon^d_{ij}\varepsilon^d_{mn}\int_{H_x} \frac{ \omega^d \xi_i \xi_j \xi_m \xi_n}{|\boldsymbol{\xi}|^2 } \mathrm{d}V_{x'} - \varepsilon^d_{ij}\varepsilon^d{ij}.
\end{split}
\end{equation}
Denote a 4th-order tensor $\mathbb{A}$ as: 
\begin{equation}
    \mathbb{A}_{ijmn}=\int_{H_x} \frac{ \omega^d \xi_i \xi_j \xi_m \xi_n}{|\boldsymbol{\xi}|^2 } \mathrm{d}V_{x'},
\end{equation}
in which $\mathbb{A}$ is a symmetrical tensor:
\begin{equation}
    \mathbb{A}_{ijmn}=\mathbb{A}_{jimn}=\mathbb{A}_{ijnm}=\mathbb{A}_{jinm}=\mathbb{A}_{mnij}=\mathbb{A}_{nmij}=\mathbb{A}_{mnji}=\mathbb{A}_{nmji}.
\end{equation}
Using this symmetrical property and  $\varepsilon^d_{ij}=\varepsilon^d_{ji}$, \cref{eq:derivOmegaD1} can be expressed as: 
\begin{equation}\label{eq:derivOmegaD2}
    \begin{split}
        0=&\varepsilon^d_{ij}\varepsilon^d_{mn} \mathbb{A}_{ijmn}- \varepsilon^d_{ij}\varepsilon^d_{ij} \\
        =&\left(\mathbb{A}_{1111} -1 \right) \varepsilon^d_{11}\varepsilon^d_{11}+\left(\mathbb{A}_{2222}-1\right) \varepsilon^d_{22}\varepsilon^d_{22}+\left(\mathbb{A}_{3333}-1\right)\varepsilon^d_{33}\varepsilon^d_{33} + \\
        & 2\mathbb{A}_{1122}\varepsilon^d_{11}\varepsilon^d_{22} + 2\mathbb{A}_{2233}\varepsilon^d_{22}\varepsilon^d_{33} + 2\mathbb{A}_{3311}\varepsilon^d_{33}\varepsilon^d_{11} + \\
        & \left(4\mathbb{A}_{1212}-2\right)\varepsilon^d_{12}\varepsilon^d_{12} + \left(4\mathbb{A}_{2323}-2\right)\varepsilon^d_{23}\varepsilon^d_{23} + \left(4\mathbb{A}_{1313}-2\right)\varepsilon^d_{13}\varepsilon^d_{13} + \\
      & 4\mathbb{A}_{1112}\varepsilon^d_{11}\varepsilon^d_{12} + 4\mathbb{A}_{1113}\varepsilon^d_{11}\varepsilon^d_{13} + 4\mathbb{A}_{2221}\varepsilon^d_{22}\varepsilon^d_{21} + 4\mathbb{A}_{2223}\varepsilon^d_{22}\varepsilon^d_{23} + 4\mathbb{A}_{3331}\varepsilon^d_{33}\varepsilon^d_{31} +
        4\mathbb{A}_{3332}\varepsilon^d_{33}\varepsilon^d_{32} + \\
    & 4\mathbb{A}_{1123}\varepsilon^d_{11}\varepsilon^d_{23} + 4\mathbb{A}_{2213}\varepsilon^d_{22}\varepsilon^d_{13} +
    4\mathbb{A}_{3312}\varepsilon^d_{33}\varepsilon^d_{12} + \\
    & 8\mathbb{A}_{1213}\varepsilon^d_{12}\varepsilon^d_{13} +                               8\mathbb{A}_{1223}\varepsilon^d_{12}\varepsilon^d_{23} +  8\mathbb{A}_{2313}\varepsilon^d_{23}\varepsilon^d_{13}.
    \end{split}
\end{equation}
At the first glance, for an arbitrary stain $\boldsymbol{\varepsilon}$, \cref{eq:derivOmegaD2} requires $\mathbb{A}_{1122}=0$ (see the 4th item) and $\mathbb{A}_{1212}=1/2$ (see the 7th item). However, from the symmetrical property of $\mathbb{A}$, we must have $\mathbb{A}_{1122}=\mathbb{A}_{1212}$. To solve this paradox, we first assume that:
\begin{equation}\label{eq:assum}
    \frac{1}{3}\mathbb{A}_{1111}=\frac{1}{3}\mathbb{A}_{2222}=\frac{1}{3}\mathbb{A}_{3333}=\mathbb{A}_{1122}=\mathbb{A}_{2233}=\mathbb{A}_{1133}.
\end{equation}
Using this assumption and $\varepsilon^d_{ii}=0$, we rearrange \cref{eq:derivOmegaD2} as:
\begin{equation}\label{eq:derivOmegaD3}
    \begin{split}
        0=&\left(\frac{2}{3}\mathbb{A}_{1111} -1 \right) \varepsilon^d_{11}\varepsilon^d_{11}+\left(\frac{2}{3}\mathbb{A}_{2222}-1\right) \varepsilon^d_{22}\varepsilon^d_{22}+\left(\frac{2}{3}\mathbb{A}_{3333}-1\right)\varepsilon^d_{33}\varepsilon^d_{33} + \\
        &\left(\frac{1}{3}\mathbb{A}_{1111} \varepsilon^d_{11}\varepsilon^d_{11} +\mathbb{A}_{1122} \varepsilon^d_{11}\varepsilon^d_{22} +\mathbb{A}_{1133} \varepsilon^d_{11}\varepsilon^d_{33} \right) +\\
         &\left(\mathbb{A}_{1122} \varepsilon^d_{11}\varepsilon^d_{22} +\frac{1}{3}\mathbb{A}_{2222} \varepsilon^d_{22}\varepsilon^d_{22} +\mathbb{A}_{2233} \varepsilon^d_{22}\varepsilon^d_{33} \right) +\\
         &\left(\mathbb{A}_{1133} \varepsilon^d_{11}\varepsilon^d_{33} +\mathbb{A}_{2233} \varepsilon^d_{22}\varepsilon^d_{33} +\frac{1}{3} \mathbb{A}_{3333} \varepsilon^d_{33}\varepsilon^d_{33} \right) +\\
        & \left(4\mathbb{A}_{1212}-2\right)\varepsilon^d_{12}\varepsilon^d_{12} + \left(4\mathbb{A}_{2323}-2\right)\varepsilon^d_{23}\varepsilon^d_{23} + \left(4\mathbb{A}_{1313}-2\right)\varepsilon^d_{13}\varepsilon^d_{13} + \\
      & 4\mathbb{A}_{1112}\varepsilon^d_{11}\varepsilon^d_{12} + 4\mathbb{A}_{1113}\varepsilon^d_{11}\varepsilon^d_{13} + 4\mathbb{A}_{2221}\varepsilon^d_{22}\varepsilon^d_{21} + 4\mathbb{A}_{2223}\varepsilon^d_{22}\varepsilon^d_{23} + 4\mathbb{A}_{3331}\varepsilon^d_{33}\varepsilon^d_{31} +
        4\mathbb{A}_{3332}\varepsilon^d_{33}\varepsilon^d_{32} + \\
    & 4\mathbb{A}_{1123}\varepsilon^d_{11}\varepsilon^d_{23} + 4\mathbb{A}_{2213}\varepsilon^d_{22}\varepsilon^d_{13} +
    4\mathbb{A}_{3312}\varepsilon^d_{33}\varepsilon^d_{12} + \\
    & 8\mathbb{A}_{1213}\varepsilon^d_{12}\varepsilon^d_{13} +                               8\mathbb{A}_{1223}\varepsilon^d_{12}\varepsilon^d_{23} +  8\mathbb{A}_{2313}\varepsilon^d_{23}\varepsilon^d_{13}\\
    =&\left(\frac{2}{3}\mathbb{A}_{1111} -1 \right) \varepsilon^d_{11}\varepsilon^d_{11}+\left(\frac{2}{3}\mathbb{A}_{2222}-1\right) \varepsilon^d_{22}\varepsilon^d_{22}+\left(\frac{2}{3}\mathbb{A}_{3333}-1\right)\varepsilon^d_{33}\varepsilon^d_{33} + \\
        & \left(4\mathbb{A}_{1212}-2\right)\varepsilon^d_{12}\varepsilon^d_{12} + \left(4\mathbb{A}_{2323}-2\right)\varepsilon^d_{23}\varepsilon^d_{23} + \left(4\mathbb{A}_{1313}-2\right)\varepsilon^d_{13}\varepsilon^d_{13} + \\
      & 4\mathbb{A}_{1112}\varepsilon^d_{11}\varepsilon^d_{12} + 4\mathbb{A}_{1113}\varepsilon^d_{11}\varepsilon^d_{13} + 4\mathbb{A}_{2221}\varepsilon^d_{22}\varepsilon^d_{21} + 4\mathbb{A}_{2223}\varepsilon^d_{22}\varepsilon^d_{23} + 4\mathbb{A}_{3331}\varepsilon^d_{33}\varepsilon^d_{31} +
        4\mathbb{A}_{3332}\varepsilon^d_{33}\varepsilon^d_{32} + \\
    & 4\mathbb{A}_{1123}\varepsilon^d_{11}\varepsilon^d_{23} + 4\mathbb{A}_{2213}\varepsilon^d_{22}\varepsilon^d_{13} +
    4\mathbb{A}_{3312}\varepsilon^d_{33}\varepsilon^d_{12} + \\
    & 8\mathbb{A}_{1213}\varepsilon^d_{12}\varepsilon^d_{13} +                               8\mathbb{A}_{1223}\varepsilon^d_{12}\varepsilon^d_{23} +  8\mathbb{A}_{2313}\varepsilon^d_{23}\varepsilon^d_{13}
    \end{split}
\end{equation}
For an arbitrary strain $\boldsymbol{\varepsilon}$, \cref{eq:derivOmegaD3} requires:
\begin{equation}\label{eq:devicondi}
    d_k=0, (k=1,2, \cdots, 15),
\end{equation}
in which $d_k$ is defined as:
\begin{align}\label{eq:odCondi}
\begin{matrix}
 d_1=\mathbb{A}_{1111}-\frac{3}{2},& d_2=\mathbb{A}_{2222}-\frac{3}{2},&  d_2=\mathbb{A}_{3333}-\frac{3}{2},\\
 d_4=\mathbb{A}_{1122}-\frac{1}{2},& d_5=\mathbb{A}_{2233}-\frac{1}{2},&  d_6=\mathbb{A}_{3311}-\frac{1}{2}, \\
 d_7=\mathbb{A}_{1112},& d_8=\mathbb{A}_{1113},& d_9=\mathbb{A}_{2221}, \\
d_{10}=\mathbb{A}_{2223},& d_{11}=\mathbb{A}_{3331},& d_{12}=\mathbb{A}_{3332}, \\
d_{13}=\mathbb{A}_{1123},& d_{14}=\mathbb{A}_{2213},&  d_{15}=\mathbb{A}_{3312}.
\end{matrix}
\end{align}
Note that \cref{eq:devicondi} satisfies and supports the assumption of \cref{eq:assum}. To find the deviatoric influence function $\omega^d$ that satisfies \cref{eq:devicondi} for arbitrary horizon shapes, we defined a function $J$ as:
\begin{equation}
    J(\omega^d,\lambda_1,\cdots, \lambda_{15})=\frac{1}{2}\left({\omega^d-\omega^d_s} \right) \bullet \left({\omega^h-\omega^h_s} \right)- \sum_{k=1}^{15} \lambda^d_k d_k,
\end{equation}
in which $\lambda^d_k$ are the Lagrange multipliers. The first variation of $J$ is:
\begin{equation}
\delta J = \left(\frac{\partial J}{\partial \omega^d } \right)^f \bullet \delta \omega^d + \sum_{k=1}^{15} \frac{\partial J}{\partial \lambda^d_k} \delta \lambda^d_k.
\end{equation}
The condition of $ \frac{\partial J}{\partial \lambda^d_k}=0$ leads to \cref{eq:devicondi}.
The condition of $\left(\frac{\partial J}{\partial \omega^d } \right)^f=0$ gives:
\begin{equation}\label{eq:omegad}
    \omega^d(\boldsymbol{\xi})=\omega^d_s(|\boldsymbol{\xi}|)+(\boldsymbol{\lambda}^d)^T \hat{\boldsymbol{\xi}}^d,
\end{equation}
in which $(\boldsymbol{\lambda}^d)^T\hat{\boldsymbol{\xi}}^d$ is the correction item for non-spherical horizons. $\boldsymbol{\lambda}^d$ and $\hat{\boldsymbol{\xi}}^d$ are defined as below:
\begin{equation}
\boldsymbol{\lambda}^d  =\left[
    \begin{matrix}
    \lambda^d_1 & \lambda^d_2 & \lambda^d_{3} & \cdots & \lambda^d_{15}
    \end{matrix}
   \right]^T, 
\end{equation}
\begin{equation}
    \begin{split}
     \hat{\boldsymbol{\xi}}^d =\frac{1}{|\boldsymbol{\xi}|^2}&\left[
    \begin{matrix}
    \xi_1^4 & \xi_2^4 & \xi_3^4 & \xi_1^2 \xi_2^2& \xi_2^2 \xi_3^2 & \xi_3^2 \xi_1^2 & \xi_1^3 \xi_2 & \xi_1^3 \xi_3 
    \end{matrix} \right.\\
    & 
    \left.\begin{matrix}
    \xi_2^3 \xi_1 & \xi_2^3 \xi_3 & \xi_3^3 \xi_1 & \xi_3^3 \xi_2 & \xi_1^2 \xi_2 \xi_3 & \xi_2^2 \xi_1 \xi_3 & \xi_3^2 \xi_1 \xi_2
    \end{matrix}
   \right]^T.
\end{split}
\end{equation}
Substituting \cref{eq:omegad} into \cref{eq:devicondi}, we have:
\begin{equation}\label{eq:lambdad}
   \boldsymbol{\lambda}^d=\left( \textbf{A}^d \right)^{-1}\textbf{R}^d,
\end{equation}
in which:
\begin{align}
    \textbf{A}^d & =\int_{H_x} \hat{\boldsymbol{\xi}}^d \otimes \hat{\boldsymbol{\xi}}^d \mathrm{d}V_{x'},\\
    \textbf{R}^d & =\textbf{r}^d - \int_{H_x} \omega_s^d \hat{\boldsymbol{\xi}}^d \mathrm{d}V_{x'},\\
\textbf{r}^d & =\left[
    \begin{matrix}
    \frac{3}{2} & \frac{3}{2} & \frac{3}{2} & \frac{1}{2} & \frac{1}{2} & \frac{1}{2} & 0 & 0 & \cdots & 0
    \end{matrix}
 \right]^T_{15 \times 1}.
\end{align}
The Lagrange multiplier $ \boldsymbol{\lambda}^d$ can be solved by \cref{eq:lambdad}. Then, the deviatoric influence function can be evaluated by \cref{eq:omegad}, which satisfies $ (\underline{\omega}^d \underline{e}^d )\bullet \underline{e}^d=\boldsymbol{\varepsilon}^d:\boldsymbol{\varepsilon}^d$ for arbitrary horizon shapes. It is worth noting that $\textbf{R}^d=0$, when $H_x$ is a sphere, we have  $\boldsymbol{\lambda}^d=0$ and $\omega^d=\omega^d_s$. Thus, the definition of the non-local SED of the XOSBPD in \cref{eq:sde_xpd} recovers that of the OSBPD.

Besides, for 2D problems, the vectors $\boldsymbol{\lambda}^d$, $\hat{\boldsymbol{\xi}}^d$ and $\textbf{r}^d$ are reduced to:
\begin{align} 
     \boldsymbol{\lambda}^d & =\left[
    \begin{matrix}
    \lambda^d_1 & \lambda^d_2 & \lambda^d_{3} & \lambda^d_{4} & \lambda^d_{5} 
    \end{matrix}
   \right]^T,\\
     \hat{\boldsymbol{\xi}}^d & = \frac{1}{|\boldsymbol{\xi}|^2}\left[
    \begin{matrix}
    \xi_1^4 & \xi_2^4 & \xi_1^2 \xi_2^2 & \xi_1^3 \xi_2 & \xi_2^3 \xi_1
    \end{matrix}
   \right]^T,\\
\textbf{r}^d & =\left[
    \begin{matrix}
    \frac{3}{2} & \frac{3}{2} & \frac{1}{2} & 0 & 0
    \end{matrix}
 \right]^T.
\end{align}

\subsection{Scalar force state}
With the derived $\omega^h$ and $\omega^d$, the non-local SED is equal to the local SED:
\begin{equation}
     W_{pd}=W_{CCM}=\frac{\kappa \theta^2}{2}+\mu(\underline{\omega}^d \underline{e}^d )\bullet \underline{e}^d,
\end{equation}
where $\mu$ is the shear modulus and $\kappa$ is the bulk modulus defined as:
\begin{equation}
\kappa=
    \begin{cases}
    \frac{E}{3(1-2\nu)}, &\text{3D}\\
    \frac{E}{2(1+\nu)(1-2\nu)}, &\text{2D plane strain}\\
    \frac{E}{2(1-\nu)}, &\text{2D plane stress}\\
    \end{cases},
\end{equation}
where $E$ is the Young's modules and $\nu$ is the Poisson's ratio.

The scalar force state $\underline{t}(\textbf{x}) \langle \boldsymbol{\xi} \rangle$ for either spherical or non-spherical horizons can be derived by using the Fréchet derivative of \cref{eq:frederi}. However, \cref{eq:frederi} requires the prerequisites (see Eqs. (90) and (95) in \cite{silling_peridynamic_2007}):
\begin{subequations}
\label{eq:prereq}
\begin{align}
    \label{eq:prereq1}
    &\left(\frac{\partial W_{pd}}{\partial \underline{e}^h} \right)^f \bullet \Delta \underline{e}^d =0,\\
    \label{eq:prereq2}
    &\left(\frac{\partial W_{pd}}{\partial \underline{e}^d} \right)^f \bullet \Delta \underline{e}^h=0.
\end{align}
\end{subequations}

Using \cref{eq:eh_xpd,eq:ed_xpd,eq:theta_xpd,eq:hrocondi,eq:devicondi} and $ \varepsilon^d_{ii} = 0$, the prerequisites of \cref{eq:prereq1,eq:prereq2} for either spherical or non-spherical horizon can be proved as:
\begin{equation}
\begin{split}
    \left(\frac{\partial W_{pd}}{\partial \underline{e}^h} \right)^f \bullet \Delta \underline{e}^d &=\kappa \theta \left(\frac{\partial \theta }{\partial \underline{e}^h}\right)^f \bullet \Delta \underline{e}^d\\
    &=\kappa \omega^h |\boldsymbol{\xi}| \theta \bullet \Delta \underline{e}^d \\
    &= \int_{H_x} \kappa \omega^h |\boldsymbol{\xi}| \theta \frac{\Delta\varepsilon^d_{ij}\xi_i\xi_j}{|\boldsymbol{\xi}|} \mathrm{d}V_{x'}\\
    &= \kappa   \theta \Delta \varepsilon^d_{ij} \int_{H_x} \omega^h \xi_i\xi_j \mathrm{d}V_{x'}\\
    &=\kappa   \theta \Delta \varepsilon^d_{ii}\\ &=0,
\end{split}
\end{equation}
and 
\begin{equation}
\begin{split}
  \left(\frac{\partial W_{pd}}{\partial \underline{e}^d} \right)^f \bullet \Delta \underline{e}^h
  &=\mu  \left(\frac{\partial \left(\underline{\omega}^d\underline{e}^d\bullet \underline{e}^d\right)}{\partial \underline{e}^d}\right)^f  \bullet \Delta \underline{e}^h \\
  &=2\mu \omega^d \underline{e}^d \bullet \Delta \underline{e}^h\\
  &= \int_{H_x} 2\mu \omega^d \frac{\varepsilon^d_{ij}\xi_i\xi_j}{|\boldsymbol{\xi}|} \frac{\Delta\varepsilon^h_{mn}\xi_m\xi_n}{|\boldsymbol{\xi}|} \mathrm{d}V_{x'}\\
    &= 2\mu \varepsilon^d_{ij} \Delta\varepsilon^h_{mn}  \int_{H_x} \omega^d \frac{\xi_i\xi_j\xi_m\xi_n}{|\boldsymbol{\xi}|^2} \mathrm{d}V_{x'}\\
    &=2\mu \varepsilon^d_{ij} \Delta\varepsilon^h_{mn} \mathbb{A}_{ijmn}\\
    &= 2\mu \varepsilon^d_{ij} \frac{\Delta\theta}{n_d} \mathbb{A}_{ijmm}\\
    &= \frac{2\mu \Delta \theta}{n_d} \frac{n_d+2}{n_d} \varepsilon^d_{ii} \\& = 0,
\end{split}
\end{equation}
in which $\theta$ is defined by \cref{eq:theta_xpd} and $\underline{e}^d$ is given by \cref{eq:ed_xpd}.

Therefore, the scalar force state $\underline{t}(\textbf{x}) \langle \boldsymbol{\xi} \rangle$ for either spherical or non-spherical horizon can be written as follows:
\begin{equation}\label{eq:t_xpd}
\begin{split}
    \underline{t}(\textbf{x}) \langle \boldsymbol{\xi} \rangle
     & =\left(\frac{\partial W_{pd}}{\partial \underline{e}^h}\right)^f + \left(\frac{\partial W_{pd}}{\partial \underline{e}^d}\right)^f \\
     &=\kappa \theta \left(\frac{\partial \theta }{\partial \underline{e}^h}\right)^f + \mu  \left(\frac{\partial \left(\underline{\omega}^d\underline{e}^d\bullet \underline{e}^d\right)}{\partial \underline{e}^d}\right)^f\\
     &=\kappa \omega^h |\boldsymbol{\xi}| \theta + 2\mu \omega^d \underline{e}^d.
\end{split}
\end{equation}
For a spherical horizon, we have $\omega^h=\omega^h_s$, $\omega^d=\omega^d_s$, and \cref{eq:t_xpd} recovers \cref{eq:finscalfor_sph}. Thus, the presented XOSBPD recovers the OSBPD when horizon is a sphere.


It is worth noting that, for some horizon shapes, $\omega^h$ and $\omega^d$ may be negative. For example, if all bonds in the horizon have $\xi_1 \xi_2>0$, there must exist at least one bond with $\omega^h<0$ because of $h_4=0$. It was stated in \citep{mitchell_position-aware_2015} that negative $\omega^h$ and $\omega^d$ are acceptable and do not introduce instability because they may not imply imaginary wave speeds. However, the authors believe that a negative $\omega^h$ is unacceptable because it violates physics. For example, the scalar force state should be negative when a body is under pure hydrostatic pressure with $\theta<0$ and $\underline{e}^d=0$. But, in this case, the evaluation of the scalar force state by \cref{eq:t_xpd} is positive if $\omega^h<0$. Furthermore, the authors found that the negative $\omega^h$ will also bring instability in the numerical implementation. To avoid this issue, the authors suggest that if there exists one bond with $\omega^h<0$ in the family, then for all bonds within this family, $\omega^h_s$ and $\omega^d_s$ have to be used to calculate the scalar force state. Note that $\omega^h$ is used to evaluate the $\theta$ to ensure its accuracy. Although this treatment introduces accuracy loss, it is acceptable because only a very few horizons have negative $\omega^h$. For example, a rectangular plate has only four horizons (centered at the four corners) with negative $\omega^h$.

Besides, for only a handful of cases, the determinants of the symmetrical matrices $\textbf{A}^h$ and $\textbf{A}^d$ may vanish. Slightly changing the horizon sizes or using a different weighted function $\omega(|\boldsymbol{\xi}|)$ may solve this problem.

\section{Numerical implementation}\label{sec:NumImp}

The static and dynamic problems can be solved with non-uniform discretization with various horizon sizes without volume and surface corrections by the XOSBPD. First, the Lagrange multipliers $\boldsymbol{\lambda}^h$ and $\boldsymbol{\lambda}^d$ are evaluated by \cref{eq:lambdah} and \cref{eq:lambdad}, respectively. Second, the influence functions $\omega^h$ and $\omega^d$ are calculated using \cref{eq:omegah} and \cref{eq:omegad}, respectively. Then, the force state is computed from \cref{eq:t_xpd,eq:T}. Finally, we compute the displacement results by the equation of motion of \cref{eq:eqm}. In this section, we briefly introduce the numerical implementation of the XOSBPD.

\subsection{Discretization}

\begin{figure}
    \centering
    \includegraphics[width=4in]{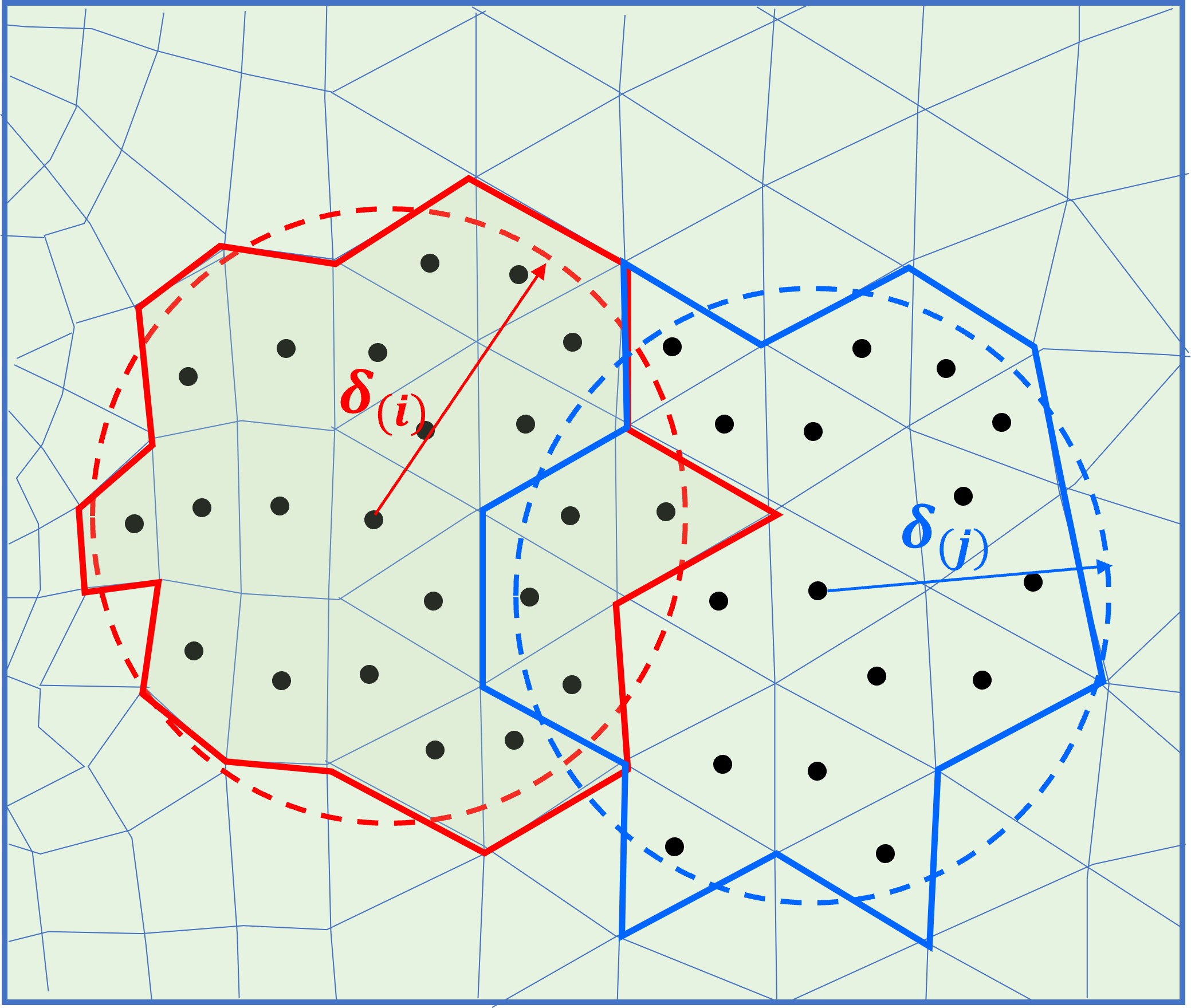}
    \caption{Numerical implementation of the XOSBPD with non-uniform discretization, non-spherical horizons and various horizon sizes. $\delta_{(i)}$ and $\delta_{j}$ are different the horizon sizes. The exact horizons are determined by \cref{eq:horizon}.}
    \label{fig:varHorizon}
\end{figure}

The XOSBPD enables the non-uniform discretization with various horizon sizes. As shown in \cref{fig:varHorizon}, the domain is discretized into non-uniform elements by mesh generators, i.e., ANSYS. The element center is the PD node $\textbf{x}_{(i)}$ associated with the element volume of $V_{(i)}$ or area of $A_{(i)}$. Note that two PD nodes $\textbf{x}_{(i)}$ and $\textbf{x}_{(j)}$ must be paired interaction for either horizon. Otherwise, it will bring ghost forces and violate the conservation of linear and angular momentum. This issue can be remedied by the following family determination:
\begin{equation}\label{eq:horizon}
    H_{\textbf{x}_{(i)}}=\left\{\textbf{x}_{(j)} \in H_{\textbf{x}_{(i)}}: \left|\textbf{x}_{(j)}-\textbf{x}_{(i)} \right| \leq \delta_{(i)} \cup \left|\textbf{x}_{(j)}-\textbf{x}_{(i)} \right| \leq \delta_{(j)} \right\},
\end{equation}
where $\delta_{(i)}=m\Delta_{(i)} $ is the horizon size, $m$ is a constant, and $\Delta_{(i)}$ is defined as:
\begin{equation}
    \Delta_{(i)}=
    \begin{cases}
    \sqrt{A_{(i)}}, \qquad \text{2-D}\\
    \sqrt[3]{V_{(i)}}, \qquad \text{3-D}
    \end{cases}.
\end{equation}
By the meshfree discretization, the one point integration \citep{silling_meshfree_2005} scheme is used for the integral of field variable $f(\textbf{x},\textbf{x}')$ as:
\begin{equation}
    \int_{H_x} f(\textbf{x},\textbf{x}') \mathrm{d}V_{x'}=\sum_{j} f(\textbf{x},\textbf{x}_{(j)})V_{(j)}.
\end{equation}
The dynamic fracture of the PD problems can be solved by the explicit scheme as follows:
\begin{subequations}
\begin{align}
 \dot{\textbf{u}}^{n+1} &= \Ddot{\textbf{u}}^n \Delta t +  \dot{\textbf{u}}^{n}, \\
    {\textbf{u}}^{n+1} &= \dot{\textbf{u}}^{n+1} \Delta t +  {\textbf{u}}^{n}. 
\end{align}
\end{subequations}
where $\Delta t$ is the time step and $n$ represents the step number.
\subsection{Adaptive dynamic relaxation}
Although the PD is essentially in a dynamic form, it also can be used to solve static or quasi-static problems using the adaptive dynamic relaxation (ADR) method \citep{Underwood_Dynamic_relaxation_1983}. The ADR approach introduces fictitious damping and inertia terms as:
\begin{equation}\label{eq:eqmADR}
    \textbf{M}\Ddot{\textbf{U}} + c \textbf{M} \dot{\textbf{U}}=\textbf{F},
\end{equation}
where \textbf{M} is the fictitious diagonal mass matrix, $\textbf{U}$ is the global displacement vector, $\textbf{F}$ is the resultant force vector, and $c$ is fictitious damping coefficient. \cref{eq:eqmADR} can be solved by the central-difference explicit method:
\begin{subequations}
\label{eq:adr}
\begin{align}
\label{eq:adr_a}
 \dot{\textbf{U}}^{n+1/2} &=\frac{(2-c^n\Delta t)\dot{\textbf{U}}^{n-1/2} + 2\Delta t \textbf{M}^{-1}
 \textbf{F}^n}{2+c^n \Delta t}, \\
 \label{eq:adr_b}
    {\textbf{U}}^{n+1} &= \dot{\textbf{U}}^{n+1/2} \Delta t +  {\textbf{U}}^{n}. 
\end{align}
\end{subequations}
where the time step incremental is generally specified as $\Delta t= 1$. The damping coefficient is given as:
\begin{equation}
c^{n}=2 \sqrt{\frac{\left(\mathbf{U}^{n}\right)^{T}{ }^{1} \mathbf{K}^{n} \mathbf{U}^{n}} {\left(\mathbf{U}^{n}\right)^{T} \mathbf{U}^{n}}},
\end{equation}
in which ${ }^{1} \mathbf{K}^{n}$ is a diagonal matrix defined as:
\begin{equation}
{ }^{1} K_{i i}^{n}=\frac{ -\left(F_{i}^{n}-F_{i}^{n-1}\right) }{M_{i i}\Delta t \dot{U}_{i}^{n-1 / 2}}, \qquad (\text{No summation for the dummy index }i),
\end{equation}
where the diagonal mass matrix is selected by:
\begin{equation}
M_{i i} \geq \frac{1}{4} \Delta t^{2} \sum_{j}\left|K_{i j}\right|, \qquad (\text{No summation for the dummy index }i),
\end{equation}
and $\textbf{K}$ is the global stiffness matrix. For the OSBPD and XOSBPD, the global stiffness matrix is not explicitly given. However, the evaluation of $\textbf{K}$ does not require good accuracy for ADR. In this study, we use the linearized bond-based PD (LBBPD) with the small displacement assumption to calculate the $\textbf{K}$ for determine the diagonal mass $M_{ii} $, because the stiffness matrix of LBBPD is very easy to construct.


%

\subsection{The global stiffness matrix of the LBBPD}

For the BBPD, the bond force is only determined by the bond but not by any other bonds:
\begin{equation}
\textbf{f}(\boldsymbol{\xi}, \boldsymbol{\eta})=CS \frac{\boldsymbol{\xi}+ \boldsymbol{\eta}}{|\boldsymbol{\xi}+ \boldsymbol{\eta}|},
\end{equation}
in which $C$ is the material constant defined as:
\begin{equation}
C=\begin{cases}
\frac{6E}{\pi \delta ^3 (1-\nu)}, \qquad &\text{2-D plane stress}\\
\frac{6E}{\pi \delta ^3 (1+\nu)(1-2\nu)}, \qquad &\text{2-D plane strain}\\
\frac{18\kappa}{\pi \delta^4}, \qquad &\text{3-D}
\end{cases},
\end{equation}
and $S$ is the stretch defined as: 
\begin{equation}
S=\frac{|\boldsymbol{\xi}+ \boldsymbol{\eta}|-|\boldsymbol{\xi}|}{|\boldsymbol{\xi}|}.
\end{equation}
Based on the small deformation assumption, the bond force can be linearized as \cite{prakash_multi-threaded_2020,zhang_ansys_2022}:
\begin{equation}
\textbf{f}(\boldsymbol{\xi}, \boldsymbol{\eta})=\frac{C}{|\boldsymbol{\xi}|^3}\left(\boldsymbol{\xi} \otimes \boldsymbol{\xi} \right) \boldsymbol{\eta}.
\end{equation}
Thus, the internal force applied on the points $\textbf{x}'$ and $\textbf{x}$ due to the bond $\boldsymbol{\xi}$ is:
\begin{equation}
\left[
    \begin{matrix}
    F(\textbf{x})\\
     F(\textbf{x}')
    \end{matrix}
    \right]=\textbf{K}_{bond}
    \left[
    \begin{matrix}
   \textbf{u} (\textbf{x})\\
     \textbf{u} (\textbf{x}')
    \end{matrix}
    \right],
\end{equation}
in which $\textbf{K}_{bond}$ is expressed as (3-D): 
\begin{equation}
    \textbf{K}_{bond}=\left[
    \begin{matrix}
    -k_{11} & -k_{12}& -k_{13} &k_{11} & k_{12}& k_{13}\\
    -k_{21} & -k_{22}& -k_{23} &k_{21} & k_{22}& k_{23}\\
    -k_{31} & -k_{32}& -k_{33} &k_{31} & k_{32}& k_{33}\\
    k_{11} & k_{12}& k_{13} &-k_{11} & -k_{12}& -k_{13} \\
    k_{21} & k_{22}& k_{23} &-k_{21} & -k_{22}& -k_{23} \\
    k_{31} & k_{32}& k_{33} &-k_{31} & -k_{32}& -k_{33} 
    \end{matrix}
    \right],
\end{equation}
where 
\begin{equation}
k_{ij}=\frac{C\xi_i \xi_j}{|\boldsymbol{\xi}|^3}\mathrm{d}V_x \mathrm{d}V_{x'}. 
\end{equation}
The global stiffness matrix of the LBBPD is assembled from the bond stiffness matrix as:
\begin{equation}
    \textbf{K}=\sum_{\text{all bonds}} \textbf{K}_{bond}.
\end{equation}

\begin{figure}[htbp]
    \centering
         \subfloat[]{\label{fig:plateHole}
    \includegraphics[width=3.2in]{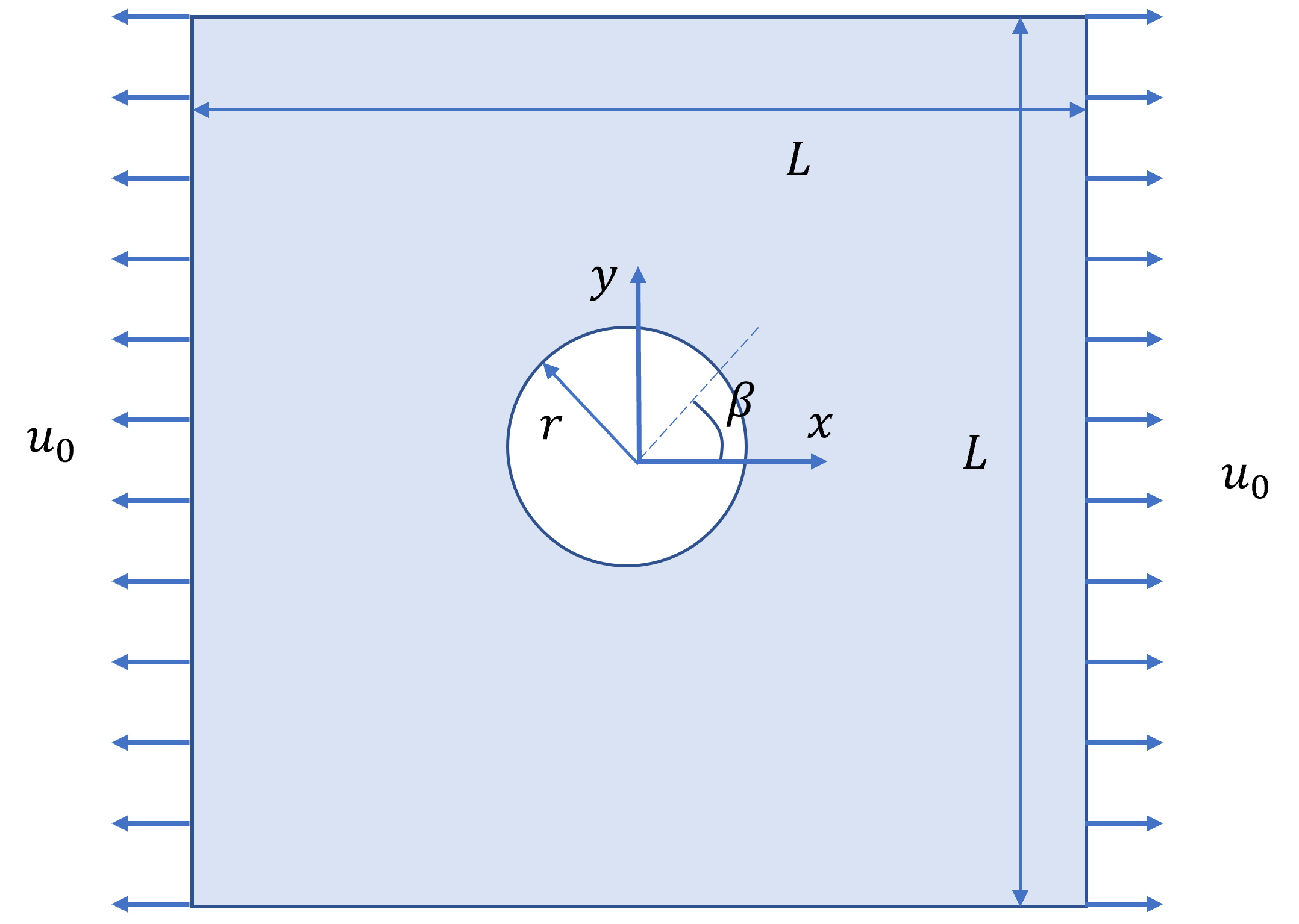}}
    \subfloat[]{\label{fig:plateHole_mesh}
    \includegraphics[width=2.3in]{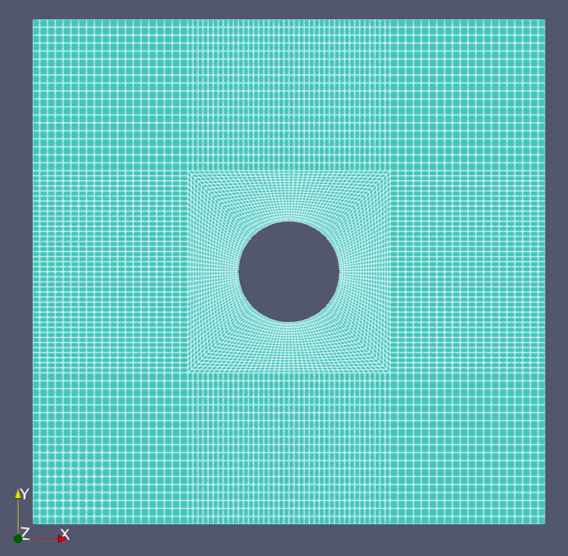}}
    \caption{A 2-D square plate with a central hole. The length of the square is $L=1$ m, the radius of the circle is $r=0.1$ m, and the displacement loading is $u_0 = 5 \times 10^{-4}$ m: (a) Geometry and loading, (b) Non-uniform meshes (7,680  non-uniform quadrilateral elements).}
    \label{fig:plate_hole_gemo_mesh}
\end{figure}

\begin{figure}[htbp]
    \centering
         \subfloat[]{\label{fig:plateHole_UX}
    \includegraphics[width=2.6in]{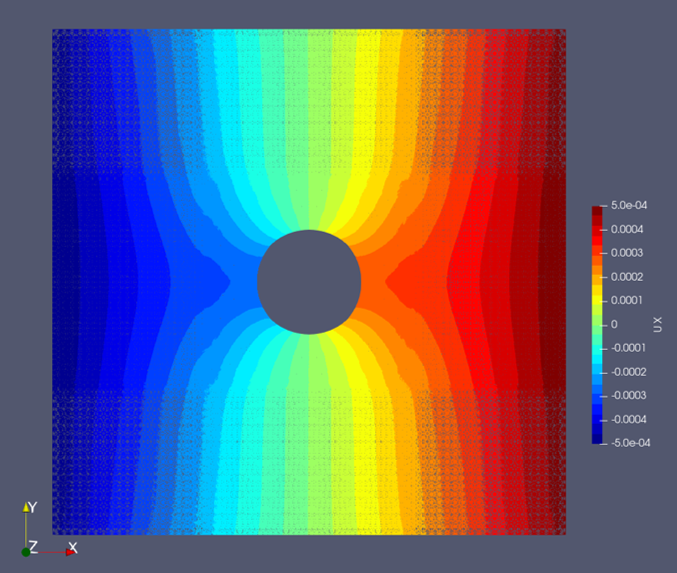}}\ \
    \subfloat[]{\label{fig:plateHole_UY}
    \includegraphics[width=2.6in]{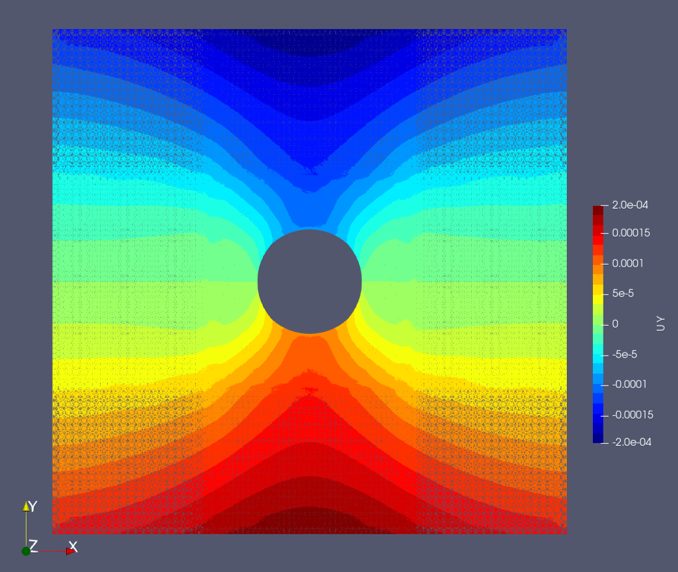}}
    \caption{The displacement results of the 2-D plate with a central hole solved by the XOSBPD with ADR (a) $u_x$, (b) $u_y$.}
    \label{fig:plateHoleU}
\end{figure}

\begin{figure}[ht!]
    \centering
         \subfloat[]{\label{fig:comPlateHoleUx}
    \includegraphics[width=2.8in]{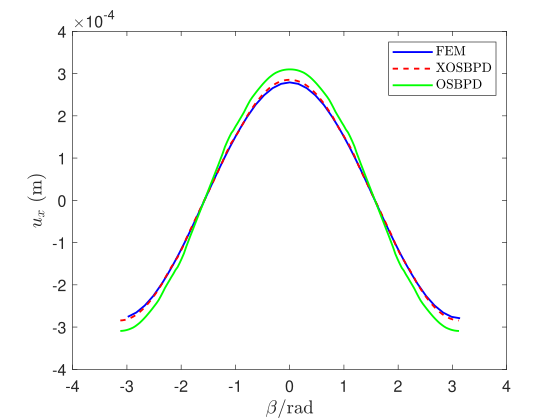}}\ \
    \subfloat[]{\label{fig:comPlateHoleUy}
    \includegraphics[width=2.8in]{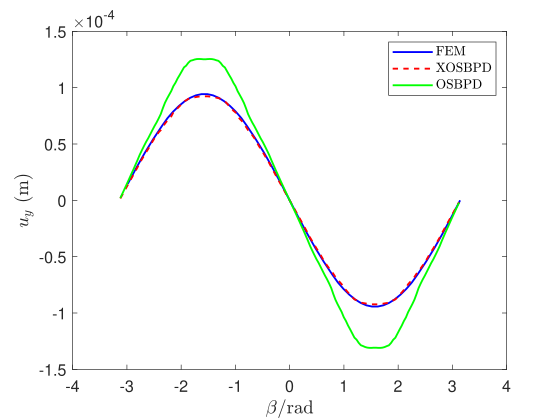}}
    \caption{The displacement results along the edge of the hole of the 2-D plate by FEM (solid blue line), XOSBPD (dotted red line), and OSBPD (green solid line): (a) $u_x$, (b) $u_y$.}
    \label{fig:CompPlateHoleU}
\end{figure}

\section{Numerical results}\label{sec:NumR}

In this section, three classic examples used in the literature are simulated and compared. The first example is a 2-D plate with a central hole under displacement loading. The second example is extruding the 2-D plate in the $z$-direction with a depth of 0.3 m. The first two examples are static problems solved by the ADR to validate the accuracy of the presented XOSBPD for both 2-D and 3-D problems. The third example is the Kalthoff-Winkler experiment, a classic study of the dynamical fracture. The Kalthoff-Winkler experiment's simulation presented in this section is used to demonstrate the XOSBPD's capability of complex dynamical fracture analysis. For all the three examples, the weighted function is specified as $\omega(|\boldsymbol{\xi}|)=1$, and the horizon is determined by $\delta_{(i)}=3.01\Delta_{(i)}$. 

\subsection{A 2-D plate with a central hole}\label{sec:plateHole}

\begin{figure}[ht!]
    \centering
         \subfloat[]{\label{fig:blockHole_UX}
    \includegraphics[width=2.in]{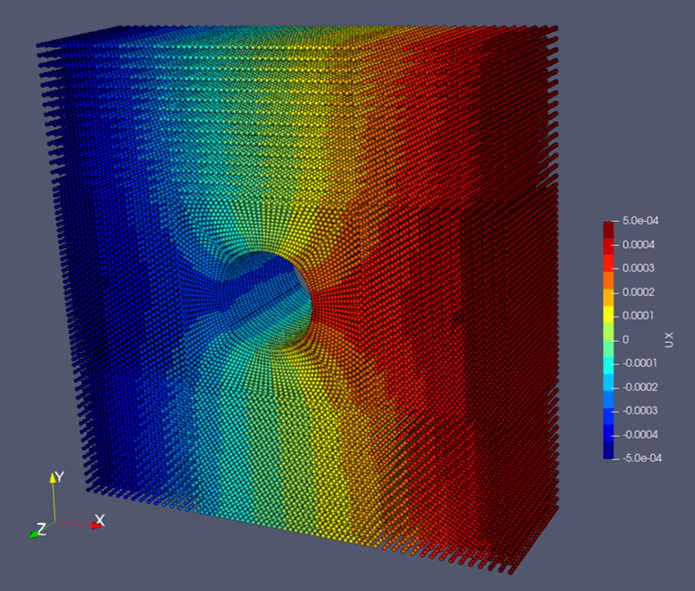}}\ 
    \subfloat[]{\label{fig:blockHole_UY}
    \includegraphics[width=2.in]{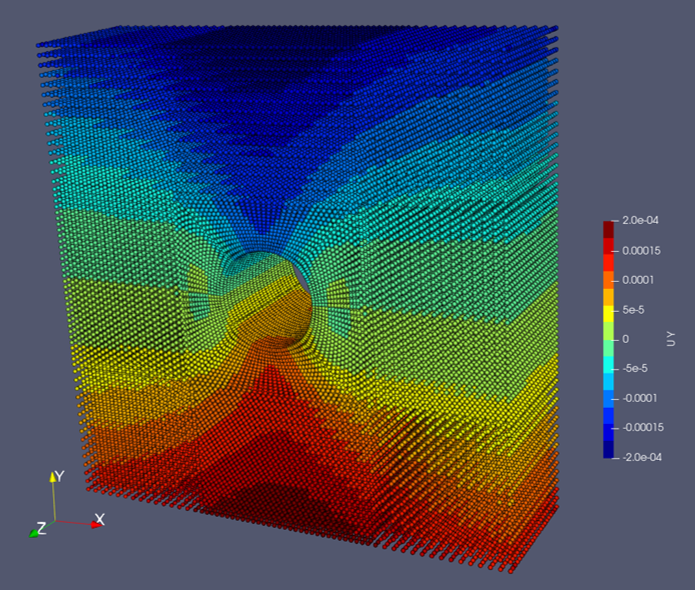}}\ 
    \subfloat[]{\label{fig:blockHole_UZ}
    \includegraphics[width=2.in]{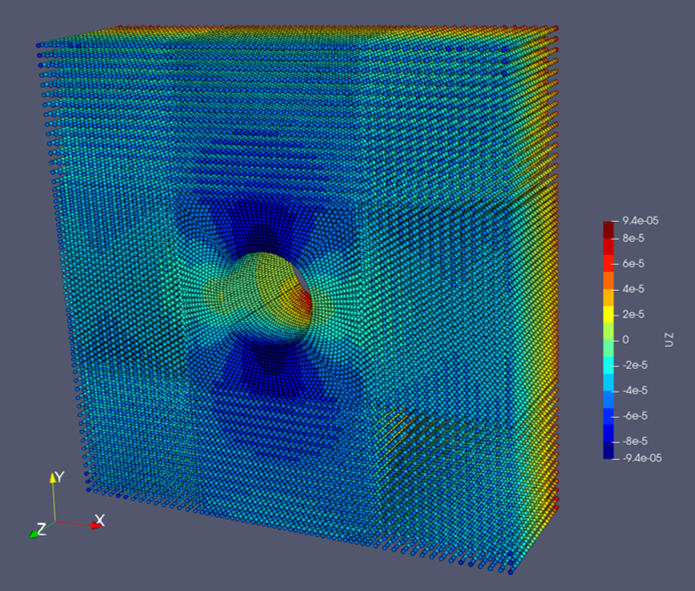}}
    \caption{The displacement results of the 3-D block with a central hole solved by the XOSBPD with ADR (a) $u_x$, (b) $u_y$, and (c) $u_z$.}
    \label{fig:blockHoleU}
\end{figure}

\begin{figure}
    \centering
         \subfloat[]{\label{fig:comBlockHoleUx}
    \includegraphics[width=2.in]{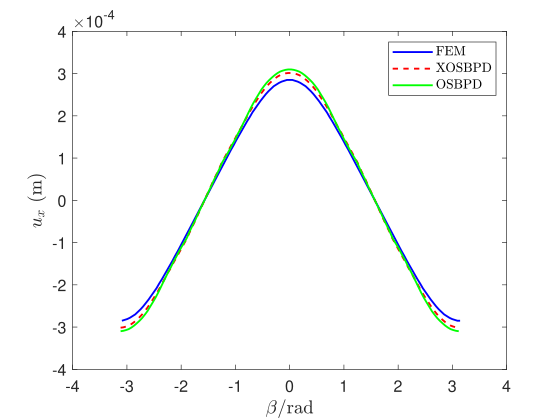}}\ 
    \subfloat[]{\label{fig:comBlockHoleUy}
    \includegraphics[width=2.in]{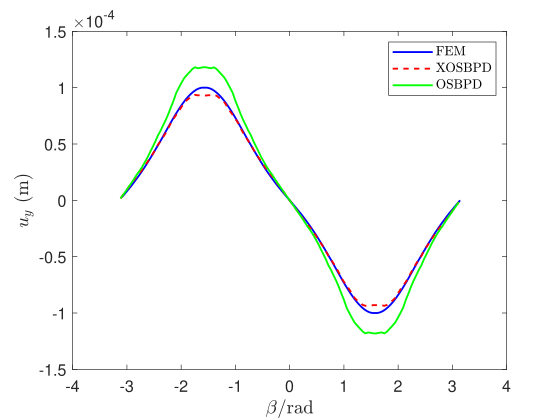}}\
    \subfloat[]{\label{fig:comBlockHoleUz}
    \includegraphics[width=2.in]{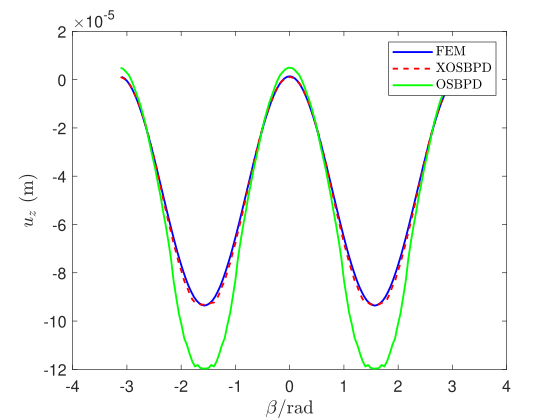}}\
    \caption{The displacement results along the edge of the hole of the 3-D block at $z=0.15$m by FEM (solid blue line), XOSBPD (dotted red line), and OSBPD (green solid line): (a) $u_x$, (b) $u_y$, (c) $u_z$.}
    \label{fig:CompBlockHoleU}
\end{figure}

\begin{figure}
    \centering
    \includegraphics[width=2.5in]{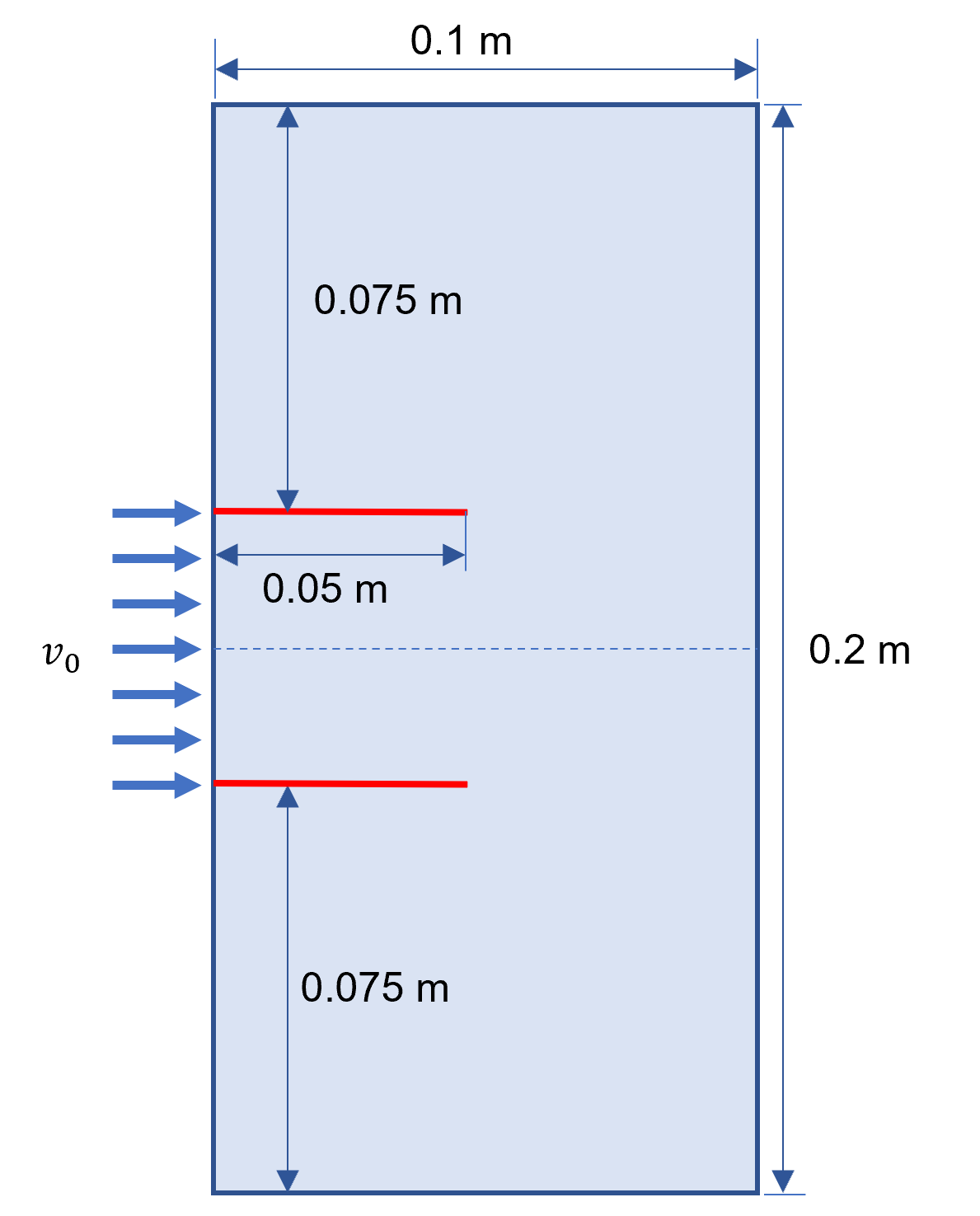}
    \caption{Geometry and loading condition of the Kalthoff-Winkler’s experiment, with red lines representing the pre-existing cracks. The 2-D plane is subject to an impact load with a speed of $v_0 = 16.5$ m/s between the two pre-existing cracks (the red lines).}
    \label{fig:KWplate}
\end{figure}

\begin{figure}[ht!]
    \centering
         \subfloat[]{\label{fig:pwPlate24us}
    \includegraphics[width=2.in]{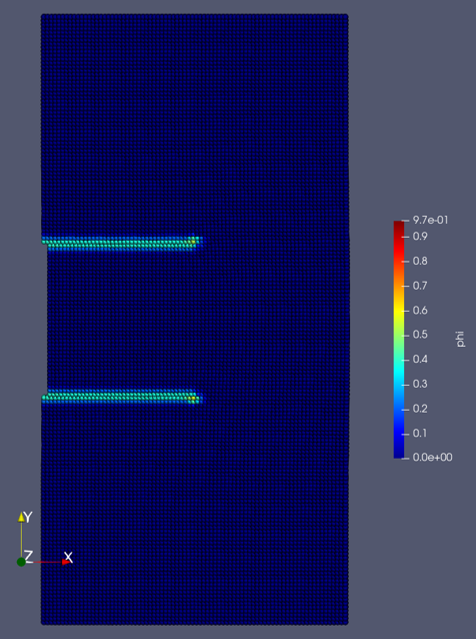}}\ 
    \subfloat[]{\label{fig:pwPlate36us}
    \includegraphics[width=2.in]{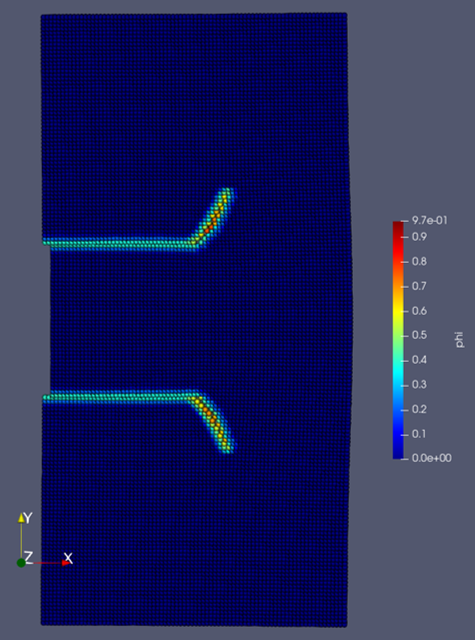}}\
    \subfloat[]{\label{fig:pwPlate48us}
    \includegraphics[width=2.in]{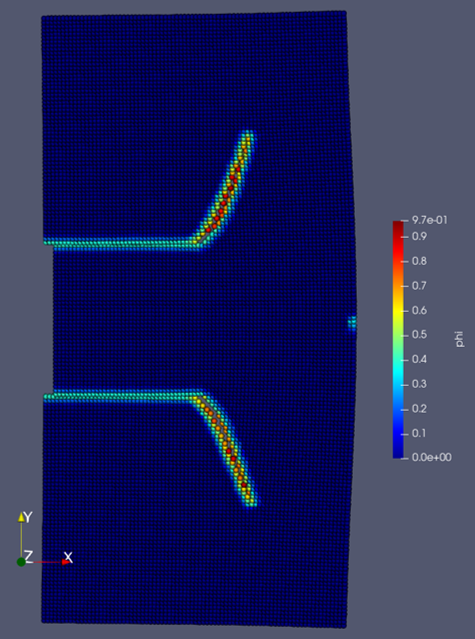}}\\
      \subfloat[]{\label{fig:pwPlate58}
    \includegraphics[width=2.in]{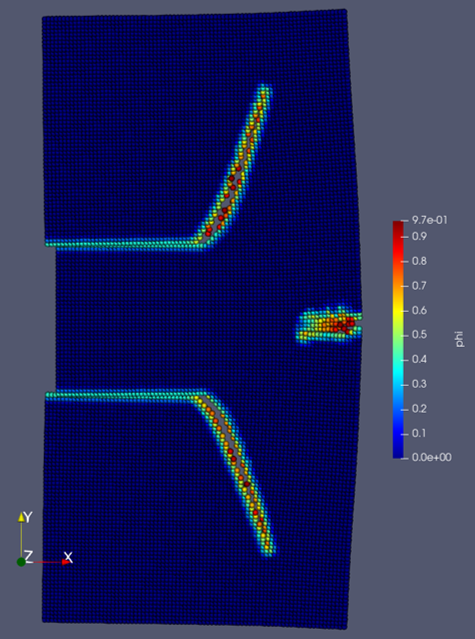}}\ 
    \subfloat[]{\label{fig:pwPlate75us}
    \includegraphics[width=2.in]{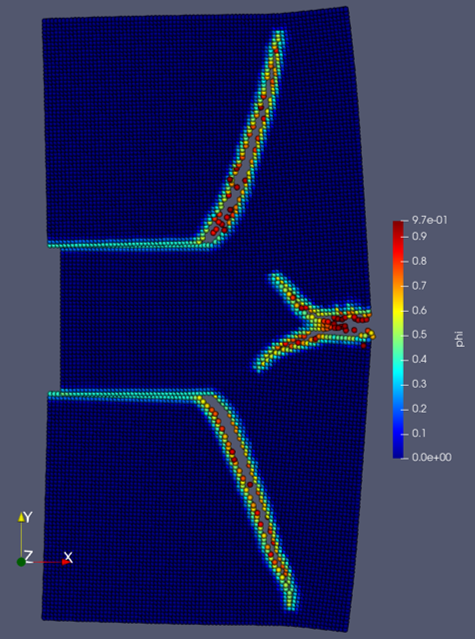}}\
    \subfloat[]{\label{fig:pwPlate104us}
    \includegraphics[width=2.in]{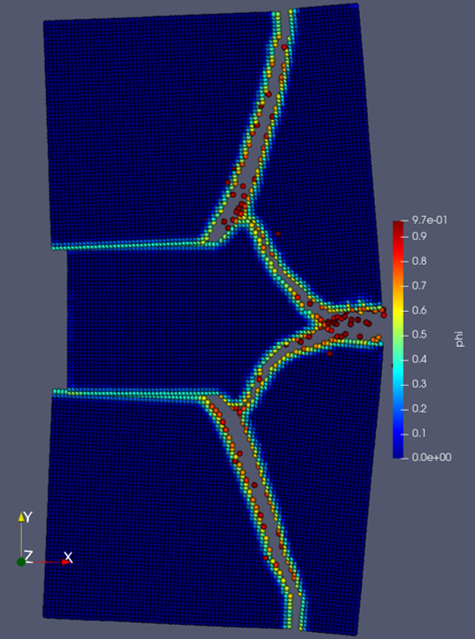}}\\
    \caption{Cracks evolution and deformation shape (scaled 5 times) of the Kalthoff-Winkler's experiment sample by the XOSBPD: (a) $t=24$ $\mu$s, th sample  of $\beta$ = 67° (respect to the $x$ axis). (b) $t=36$ $\mu$s, the two new cracks are growing in a linear manner. (c) $t=48$ $\mu$s, new damage emerges close to the center of the right-hand surface. (d) $t=58.4$ $\mu$s, the new damage branches into two cracks. (e) $t=75$ $\mu$s,  the two new cracks keep growing. and (f) $t=104$ $\mu$s, the two pre-existing and the new two cracks meet each other and the whole plane breaks into five pieces.}
    \label{fig:pwPlateCrack}
\end{figure}

\begin{figure}[ht!]
    \centering
     \subfloat[]{\label{fig:wp1DEM}
    \includegraphics[width=3.in]{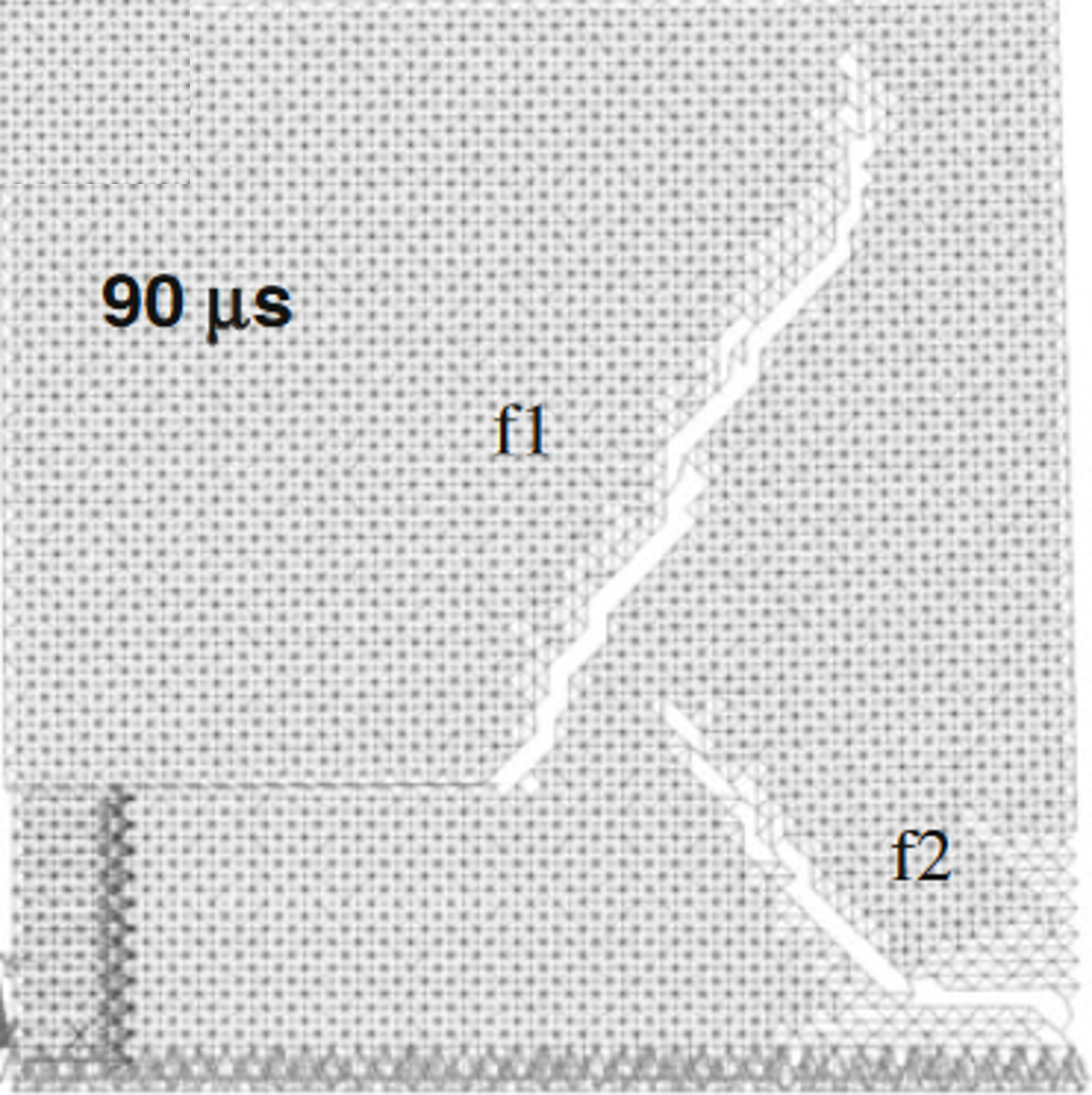}}\  \
    \subfloat[]{\label{fig:wp1XFEM}
    \includegraphics[width=3.in]{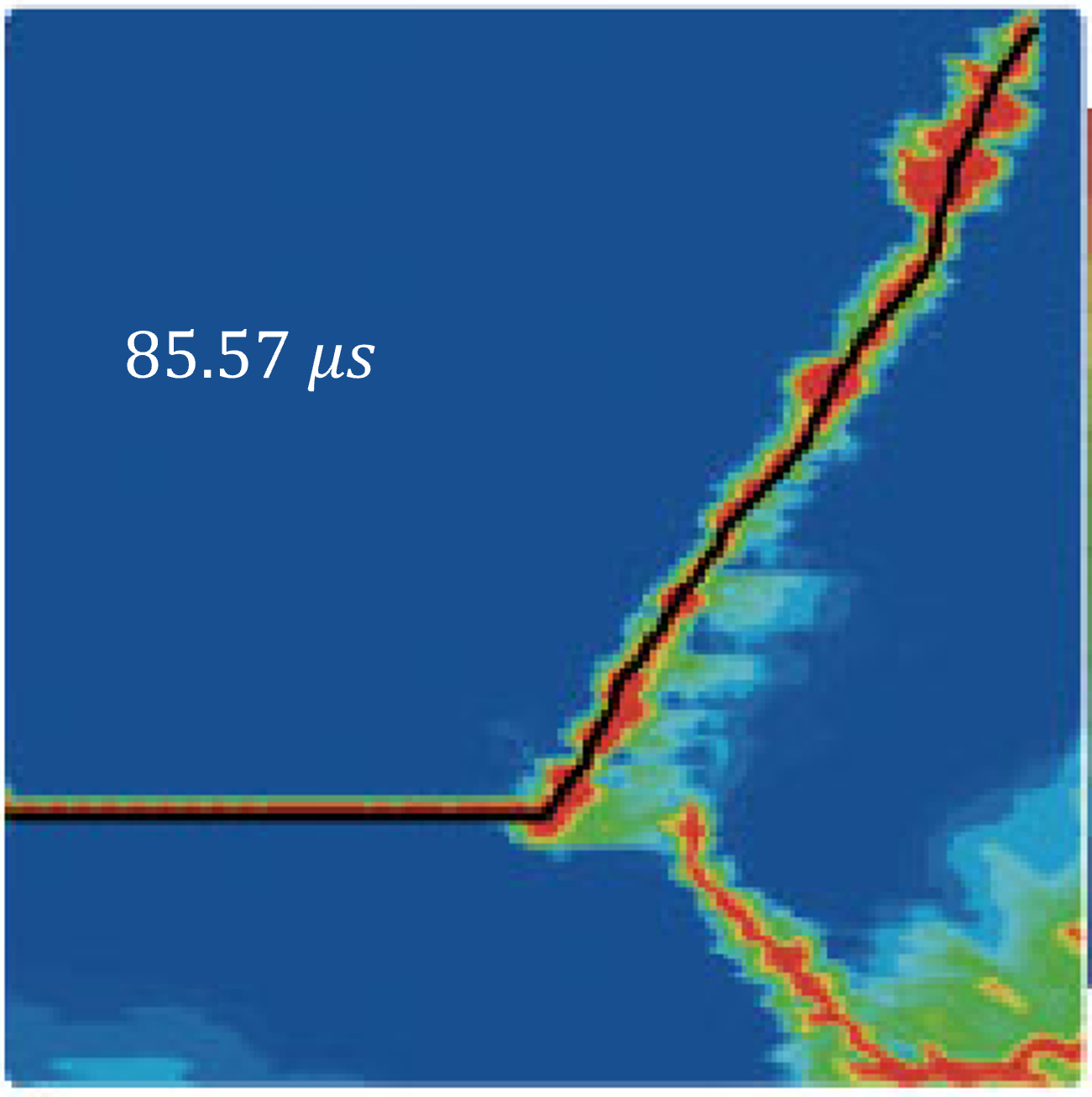}}\
    \caption{Cracks propagation simulation of the Kalthoff-Winkler's experiment using (a) DEM \cite{kosteski_crack_2012} and (b) XFEM \cite{belytschko_dynamic_2003}.}
    \label{fig:pwPlateCrackLite}
\end{figure}
As shown in Fig. \ref{fig:plateHole}, a 2-D plane stress plate with a central hole is subject to a displacement loading of $ u_0=5\times 10^{-4}\ \mathrm{m}$. The dimensions of the plate are $L=1$ m and $r=0.1$ m. The material properties of the plate are Young's modulus $E=70\ \mathrm{GPa}$ and Poisson's ratio $\nu=0.33$. The whole model is discretized into 7,680 non-uniform quadrilateral elements, and thus with a total of 7,680 PD nodes, as illustrated in \cref{fig:plateHole_mesh}.


\cref{fig:plateHoleU} illustrates the displacement results of this example solved by the XOSBPD with the ADR technique. We also simulate this example using the original OSBPD without volume and surface corrections and compare the results with FEM. \cref{fig:CompPlateHoleU} shows the displacement comparison results by the three methods along the edge of the hole. The arc angle $\beta$ along the edge is defined in \cref{fig:plateHole}. As shown in \cref{fig:CompPlateHoleU}, for both the horizontal displacement $u_x$ and the vertical displacement $u_y$, the XOSBPD is very close to the FEM while the OSBPD has a larger error. The displacement results of this 2-D example confirm that the presented XOSBPD works for the non-spherical horizon and non-uniform mesh. 

\subsection{A 3-D block with a central hole}\label{sec:blockHole}

After the accuracy demonstration of the 2-D case of the XOSBPD, We extrude the 2-D plate of the first example in the $z$-direction with a depth of 0.3 m and simulate the 3-D block with the same loading conditions and same material properties as the first example. The 3-D block is discretized into 173,944 hexahedron elements, with a total number of 173,944 PD nodes. 

\cref{fig:plateHoleU} illustrates domain discretization and the displacement results of this example solved by the presented XOSBPD with ADR technique.


Similar to the first example, \cref{fig:CompBlockHoleU} shows the displacement results by the FEM, XOSBPD, and original OSBPD methods along the edge of the hole on the surface $z=0.15$ m. As we can see, the displacement results of XOSBPD are very close to that of FEM. But for OSBPD, the results have significant errors, especially for $u_y$ and $u_z$. The displacement results of this 3-D example further confirm that the presented XOSBPD works for the non-spherical horizon and non-uniform mesh of 3-D problems.

\subsection{The Kalthoff-Winkler’s experiment} \label{sec:KW}

This section compares the dynamical fracture simulation of the well-known Kalthoff-Winkler's experiment \citep{kalthoff_modes_nodate_2000}. As depicted in \cref{fig:KWplate}, a 2-D plane strain plate is imposed by impact load with a speed of $v_0=16.5$ m/s between the two pre-existing cracks. The other boundaries are free. The material properties are: the Young's modulus $E=190$ GPa, the mass density $\rho=8,000\ \mathrm{kg/m^3}$, the Poisson's ratio $\nu=0.25$, and the critical fracture energy release rate $G_0=222,170\ \mathrm{J/m^2}$. A uniform mesh discretizes the domain with a size of 1.25 mm. The time step is specified as $\Delta t=80$ ns. For this example, the bond stretch criterion is used. That is, if the bond stretch $S$ exceeds the critical value $S_0$, the bond will be irreversibly broken. The critical stretch for plane strain is defined as \cite{zaccariotto_luongo_sarego_galvanetto_2015}:
 \begin{equation}
     S_0=\sqrt{\frac{5\pi G_0}{12E\delta}}.
 \end{equation}
 
The simulation results are presented in \cref{fig:pwPlateCrack}. As we can see, the two pre-existing cracks start to grow from around $t=24\ \mathrm{\mu s}$ with the inclined angle of 67° (respect to the $x$ axis), which is very close to 70° from the experimental observation \citep{kalthoff_modes_nodate_2000}. At around time $t=48\ \mathrm{\mu s}$, new damage emerges close to the center of the right-hand surface. At time $t=58.4\ \mathrm{\mu s}$, the new crack branches into two cracks and finally coalesce with the two old cracks (see \cref{fig:pwPlate104us}). The cracks growth trigger time ($t=24\ \mathrm{\mu s}$) is also obtain by \citet{kosteski_crack_2012}  using discrete element method (DEM) and by \citet{belytschko_dynamic_2003} using extended FEM (XFEM), and they also observed the new crack and its braching, as shown in \cref{fig:pwPlateCrackLite}.

\section{Conclusion}\label{sec:conc}


This work extends the OSBPD model from the spherical horizon to non-spherical ones. The formulation is achieved by introducing the Lagrange multipliers to ensure the non-local dilatation and non-local SED equal the local dilatation and local SED, respectively. Since the XOSBPD works for arbitrary horizon shapes, volume and surface corrections are no longer needed. Moreover, the non-uniform discretization implementation with various horizon sizes is also made possible by the XOSBPD, which can save computation costs and are more conformal to complex geometries. The first two static examples solved by the ADR show that the XOSBPD has excellent accuracy for both 2-D and 3-D problems. The third example compares with the Kalthoff-Winkler experiment and confirms the XOSBPD's complex dynamical fracture analysis capability. The proposed method paves a road to the comprehensive study of static and dynamics of the failure mechanism of structure components and solid materials.


\bibliographystyle{elsarticle-num-names} 

\bibliography{cas-refs}

\end{document}